\documentclass[aps,prc,superscriptaddress,floats,floatfix,reprint,reprintnumbers]{revtex4}
\usepackage{blindtext}
\pdfoutput=1
\usepackage{epsfig}
\usepackage{graphicx}
\usepackage{dcolumn}
\usepackage{bm}
\usepackage{booktabs}
\usepackage{color}
\usepackage{mathrsfs}
\usepackage{ulem}
\usepackage{epstopdf}
\graphicspath{{plots/}}
\usepackage{verbatim}
\usepackage{indentfirst}
\usepackage{float}
\usepackage[colorlinks,
           bookmarks=true,
           linkcolor=blue,
           urlcolor=blue,
            anchorcolor=black,
            citecolor=blue]{hyperref}
\usepackage[bf,raggedright,indentafter]{titlesec}
\usepackage{rotating}
\usepackage{booktabs}
\usepackage{tabularx}
\usepackage{makecell}
\usepackage{hypcap}
\usepackage{subfigure}%
\usepackage{tabularx}

\usepackage{array}

\begin{document}
\hspace{3cm}
\title{\Large Four-body  Semileptonic Charm Decays  $D\to P_1P_2\ell^+\nu_\ell $  Based on \\  SU(3) Flavor Analysis}
\author{Ru-Min Wang$^{1,\dagger}$\thanks{ruminwang@sina.com},~~~Yi Qiao$^{1}$,~~~Yi-Jie Zhang$^{1}$,~~~Xiao-Dong Cheng$^{2,\S}$,~~~Yuan-Guo Xu$^{1,\sharp}$\\
 $^1${\scriptsize College of Physics and Communication Electronics, Jiangxi Normal University, Nanchang, Jiangxi 330022, China}\\
 $^2${\scriptsize College of Physics and Electronic Engineering, Xinyang Normal University, Xinyang, Henan 464000, China}\\
  $^\dagger${\scriptsize ruminwang@sina.com}~~
 $^\S${\scriptsize chengxd@mails.ccnu.edu.cn}~~
 $^\sharp${\scriptsize yuanguoxu@jxnu.edu.cn}~~
 }

\vspace{3cm}

\begin{abstract}
Motivated by the significant experimental progress in probing  semileptonic decays  $D\to P_1P_2\ell^+\nu_\ell~(\ell=\mu,e)$,
we analyze the branching ratios of the $D\to P_1P_2\ell^+\nu_\ell $ decays with the nonresonant, the light scalar meson resonant and the vector meson resonant contributions in this work.
We obtain  the hadronic amplitude relations   between different decay modes by the SU(3) flavor analysis, and then predict relevant branching ratios of the $D\to P_1P_2\ell^+\nu_\ell$ decays by  the present experimental data with $2 \sigma$ errors.  Most of our predicted  branching ratios are consistent with present experimental data within $2\sigma$ error bars, and others are consistent with the data within $3\sigma$ error bars.
We find that the branching ratios of the nonresonant decays $D^0\to \pi^-\overline{K}^0\ell^+\nu_\ell,\pi^0K^-\ell^+\nu_\ell$, $D^+\to \pi^+K^-\ell^+\nu_\ell,\pi^0\overline{K}^0\ell^+\nu_\ell,\pi^+\pi^-\ell^+\nu_\ell,\pi^0\pi^0\ell^+\nu_\ell$, and $D^+_s\to K^+K^-\ell^+\nu_\ell,K^0\overline{K}^0\ell^+\nu_\ell$   are on the order of $\mathcal{O}(10^{-3}-10^{-4})$. %
%
The vector meson resonant contributions are dominant in the $D^0\to \pi^-\overline{K}^0\ell^+\nu_\ell,\pi^0K^-\ell^+\nu_\ell,\pi^0\pi^-\ell^+\nu_\ell$, $D^+\to \pi^+K^-\ell^+\nu_\ell,\pi^0\overline{K}^0\ell^+\nu_\ell,\pi^+\pi^-\ell^+\nu_\ell$, and $D^+_s\to K^+K^-\ell^+\nu_\ell,K^0\overline{K}^0\ell^+\nu_\ell, K^+\pi^-\ell^+\nu_\ell, K^0\pi^0\ell^+\nu_\ell$ decays.
The nonresonant, the vector meson resonant  and the scalar resonant contributions are all important in the $D^0\to\eta\pi^-\ell^+\nu_\ell$ decays.
The  $D^0\to K^-K^0\ell^+\nu_\ell,\eta'\pi^-\ell^+\nu_\ell$ and $D^+\to \overline{K}^0K^0\ell^+\nu_\ell,\pi^0\pi^0\ell^+\nu_\ell,\eta\pi^0\ell^+\nu_\ell,\eta'\pi^0\ell^+\nu_\ell$ decays only receive  both  the nonresonant and the scalar resonant contributions, and both contributions  are  important in their branching ratios.
According to our predictions, many decay modes could be observed in the experiments like  BESIII, LHCb, and BelleII, and some decay modes  might be measured in these experiments in the near future.

\end{abstract}
\maketitle

\newpage
\section{INTRODUCTION}
Semileptonic heavy meson decays  dominated by tree-level exchange of $W$-bosons in the SM
are very important
processes in testing the standard model and in
searching for the new physics beyond the standard model, for
example, the extraction of the Cabbibo-Kobayashi-Maskawa (CKM) matrix elements.
Four-body semileptonic exclusive decays $D\to P_1P_2\ell^+\nu_\ell$ are generated by the  $c\to s/d\ell^+\nu_\ell$ transitions,
and they can  receive contributions from the nonresonant, the light scalar meson resonant and the vector meson resonant contributions, etc.
Therefore, these decays are also a good laboratory for probing the internal  structure of light  hadrons \cite{Wang:2009azc,Oset:2016lyh,Achasov:2012kk}.
Some  nonresonant $D\to P_1P_2\ell^+\nu_\ell$ decays, the light scalar meson resonant decays $D\to S (S\to P_1P_2) \ell^+\nu_\ell$  and the vector meson resonant decays $D\to S (S\to P_1P_2) \ell^+\nu_\ell$
have been observed  by  BESIII, BABAR, CLEO, and MARKIII \cite{PDG2022,BESIII:2018qmf,BESIII:2018sjg,BESIII:2021tfk,BESIII:2021pdt,BaBar:2010vmf,CLEO:2009ugx,MARK-III:1990bbt}.  Present experimental measurements  give us an opportunity to  additionally test  theoretical approaches.

Experimental backgrounds of  the semileptonic decays are cleaner than ones of the hadronic decays, and
theoretical description of the semileptonic exclusive decays are relatively simple.  Since leptons do not participate in the strong interaction, the weak and strong dynamics
can be separated in these processes. All the strong dynamics in
the initial and final hadrons is included in the hadronic transition form factors, which are important for testing the theoretical calculations of the involved strong interaction.
The form factors  can be calculated, for example, by the chiral perturbation theory \cite{Kang:2013jaa}, the unitarized chiral perturbation theory \cite{Shi:2017pgh,Shi:2020rkz},
the light-cone sum rules \cite{Sekihara:2015iha,Cheng:2017smj,Hambrock:2015aor},  and the QCD factorization \cite{Boer:2016iez}. Nevertheless, due to our poor understanding
of hadronic interactions, the evaluations of the form factors  are difficult and   often plugged with large uncertainties.  One needs to find ways to minimize the uncertainties to
extract useful information.

In the lack of reliable calculations, symmetries provide very important information for particle physics. SU(3) flavor symmetry is a symmetry
in QCD for strong interaction.
From the perspective of the SU(3) flavor symmetry,  the leptonic part of the $D\to P_1P_2\ell^+\nu_\ell$ decay is the SU(3) flavor singlet,  which  makes
no difference between different decay modes with certain lepton ($e$ or $\mu$).
The different hadronic parts (the hadronic amplitudes or the hadronic  form factors) of the $D\to P_1P_2\ell^+\nu_\ell$ decays could be related by the SU(3) flavor symmetry without the detailed dynamics.
Nevertheless, the size of the hadronic amplitudes or the form factors cannot be determined by itself in the SU(3) flavor symmetry approach.
However, if experimental data are enough, one may use the data to extract the hadronic amplitudes or the form factors, which can be viewed
as predictions based on symmetry,  and  has a smaller dependency on estimated form
factors. Although the SU(3) flavor symmetry is only an approximate symmetry because up, down
and strange quarks have different masses, it still provides some very useful information about the
decays. The SU(3) flavor symmetry  has been widely used to study hadron decays, for instance,  $b$-hadron decays  \cite{He:1998rq,He:2000ys,Fu:2003fy,Hsiao:2015iiu,He:2015fwa,He:2015fsa,Deshpande:1994ii,Gronau:1994rj,Gronau:1995hm,Shivashankara:2015cta,Zhou:2016jkv,Cheng:2014rfa,Wang:2021uzi,Wang:2020wxn},   $c$-hadron decays  \cite{Wang:2021uzi,Wang:2020wxn,Grossman:2012ry,Pirtskhalava:2011va,Savage:1989qr,Savage:1991wu,Altarelli:1975ye,Lu:2016ogy,Geng:2017esc,Geng:2018plk,Geng:2017mxn,Geng:2019bfz,Wang:2017azm,Wang:2019dls,Wang:2017gxe,Muller:2015lua}, and light hadron decays \cite{Wang:2019alu,Wang:2021uzi,Xu:2020jfr,Chang:2014iba,Zenczykowski:2005cs,Zenczykowski:2006se,Cabibbo:1963yz}.

Although the SU(3) flavor symmetry works well in  heavy hadron decays, the calculations of SU(3) flavor breaking effects   would play a key role in the  precise theoretical predictions of the observables and a precise test of the unitarity of the CKM matrix.  If up and down quark masses are neglected, a
nonzero strange quark mass breaks the SU(3) flavor  symmetry down to the isospin symmetry.
When up and down quark mass difference is kept, isospin symmetry is also broken.
Applications
of the SU(3) flavor breaking approach  on hadron decays can be found in Refs. \cite{Dery:2020lbc,Sasaki:2008ha,Pham:2012db,Geng:2018bow,Flores-Mendieta:1998tfv,Cheng:2012xb,Xu:2013dta,He:2014xha}.
The SU(3) flavor breaking effects due to the fact of $m_s\gg m_{u,d}$ will be considered in our  analysis of the nonresonant $D\to P_1P_2\ell^+\nu_\ell$ decays.

Four-body semileptonic decays $D\to P_1P_2\ell^+\nu_\ell$ have been studied, for instance, in Refs. \cite{Shi:2021bvy,Shi:2017pgh,Kim:2017dfr,Achasov:2020qfx,Wiss:2007mr,Wang:2016wpc,Achasov:2021dvt}.
In this work,  we will  study the $D\to P_1P_2\ell^+\nu_\ell$ decays with the SU(3) flavor symmetry/breaking. In three cases of  the nonresonant decays, the light scalar meson resonant decays and the vector meson resonant decays,  we  will  firstly construct the hadronic amplitude relations   between different decay modes,
use the available data to extract the hadronic amplitudes,   then predict the  not-yet-measured modes for further tests in experiments, and finally  analyze the  contributions with the non-resonance, the light scalar meson resonances and the vector meson resonances in the  branching ratios.

This paper is organized as follows. In Sec. II, the expressions of the branching ratios are given. In Sec. III, we will give our numerical results of the $D\to P_1P_2\ell^+\nu$ decays with the nonresonant, the light scalar meson resonant and the vector meson resonant contributions.  Our conclusions are given in Sec. IV.

\section{Theoretical frame}
\subsection{Decay branching ratios}
The effective Hamiltonian for $c\to q_i\ell^+\nu_\ell$ transition can be written as
\begin{eqnarray}
\mathcal{H}_{eff}(c\rightarrow q_i\ell^+\nu_\ell)&=&\frac{G_F}{\sqrt{2}}V_{cq_i}\bar{q}_i\gamma^\mu(1-\gamma_5)c~\bar{\nu}_\ell\gamma_\mu(1-\gamma_5)\ell,\label{Heff}
\end{eqnarray}
where $G_F$ is the Fermi constant,   $V_{cq_i}$  is the CKM matrix element, and $q_i=d,s$ for $i=2,3$.
The decay amplitude of the $D(p)\to P_1(k_1)P_2(k_2)\ell^+(q_1)\nu_\ell(q_2)$ decay can be divided into  leptonic and hadronic parts
\begin{eqnarray}
\mathcal{A}(D\rightarrow P_1P_2\ell^+\nu_\ell)&=&\langle P_1(k_1)P_2(k_2)\ell^+(q_1)\nu_\ell(q_2)| \mathcal{H}_{eff}(c\rightarrow q_i\ell^+\nu_\ell) |D(p)\rangle\\
&=&\frac{G_F}{\sqrt{2}}V_{cq_i} L_{\mu}H^{\mu},
\end{eqnarray}
where  $L_\mu=\bar{\nu_\ell}\gamma_{\mu}(1-\gamma_5)\ell$ is the leptonic  charged current, and $H^\mu=\langle P_1(k_1)P_2(k_2)|\bar{s}/\bar{d}\gamma^\mu(1-\gamma_5)c|D(p)\rangle$ is the hadronic matrix element.  The leptonic part $L_\mu$  is calculable using the
perturbation theory, while the hadronic part $H^\mu$ is encoded into the transition form factors.  Following  Refs. \cite{Boer:2016iez,Faller:2013dwa},  the $D\to P_1P_2$ form factors are given as
\begin{eqnarray}
\langle P_1(k_1)P_2(k_2)|\bar{s}/\bar{d}\gamma^\mu c|D(p)\rangle&=&iF_\perp \frac{1}{\sqrt{k^2}}q^\mu_{\perp},  \\
-\langle P_1(k_1)P_2(k_2)|\bar{s}/\bar{d}\gamma^\mu\gamma_5c|D(p)\rangle&=&F_t \frac{q^\mu}{\sqrt{q^2}}+F_0 \frac{2\sqrt{q^2}}{\sqrt{\lambda}}k^\mu_{0}+F_\parallel \frac{1}{\sqrt{k^2}}\bar{k}^\mu_{\parallel}, \label{Eq:DFFB2PP}
\end{eqnarray}
with
\begin{eqnarray}
k^\mu_{0}&=&k^\mu-\frac{k\cdot q}{q^2}q^\mu, \\
\bar{k}^\mu_{\parallel}&=&\bar{k}^\mu-\frac{4(k\cdot q)(q\cdot \bar{k})}{\lambda}k^\mu+\frac{4k^2(q\cdot \bar{k})}{\lambda}q^\mu,\\
q^\mu_{\perp}&=& 2\epsilon^{\mu\alpha\beta\gamma}\frac{q_\alpha k_\beta \bar{k}_\gamma}{\sqrt{\lambda}},
\end{eqnarray}
where $k\equiv k_1+k_2$, $q\equiv q_1+q_2$, $\bar{k}\equiv k_1-k_2$, $\bar{q}\equiv q_2-q_1$, and $\lambda=\lambda(m_D^2,q^2,k^2)$ with $\lambda(a,b,c)= a^2+b^2+c^2-2ab-2bc-2ac$.

In terms of the form factors,   the differential branching ratio  of the nonresonant $D\to P_1P_2\ell^+\nu_\ell$ decays can be written as \cite{Boer:2016iez}
\begin{eqnarray}
\frac{d\mathcal{B}(D\to P_1P_2\ell^+\nu)_N}{dq^2~dk^2}=\frac{1}{2}\tau_{D}|\mathcal{N}|^2\beta_\ell(3-\beta_\ell)|F_A|^2, \label{Eq:DB2PPlvdbr}
\end{eqnarray}
with
 \begin{eqnarray}
|\mathcal{N}|^2&=&G_F^2|V_{cq}|^2\frac{\beta_\ell q^2\sqrt{\lambda(m_D^2,q^2,k^2)}}{3\cdot2^{10}\pi^5m_D^3}~~~~\mbox{with}~~~~\beta_\ell=1-\frac{m_\ell^2}{q^2},  \nonumber \\
|F_A|^2&=& |F_0|^2+\frac{2}{3}(|F_{\parallel}|^2+|F_\perp|^2)+\frac{3m_\ell^2}{q^2(3-\beta_\ell)}|F_t|^2,\label{Eq:DB2PPlvdbr}
\end{eqnarray}
where $\tau_{M}$($m_{M}$) is lifetime(mass) of $M$ particle. In this work, we ignore the small contributions  of the $|F_t|^2$ term, which is  proportional to $m_\ell^2$. The corresponding limits of integration are given by  $(m_{P_1}+m_{P_2})^2\leq k^2\leq(m_{D_q}-m_\ell)^2$ and $m_\ell^2\leq q^2 \leq(m_{D_q}-\sqrt{k^2})^2$.
 The calculations of the form factors $F_0,~F_{\parallel},~F_\perp$, and $F_t$  are quite complicated, and their specific expressions in the QCD factorization limit can be found in Ref. \cite{Boer:2016iez}.  Nevertheless, we will not use the specific expressions in this work, and we will relate  the different hadronic decay amplitudes or the different  form factors  between different decay modes by the SU(3) flavor symmetry/breaking, which are discussed  in later Sec. \ref{sec:FFSU3orB}.

Except for the nonresonant $D\rightarrow P_1P_2\ell^+\nu_\ell$ decays, the resonant $D\to R(R\to P_1P_2)\ell^+\nu_\ell$  decays with the scalar($R=S$) resonance  and the vector($R=V$) resonance are also studied in this work.
In the case of  the decay widths of the  resonances are  very narrow, the
resonant decay branching ratios  respect  a simple factorization relation
 \begin{eqnarray}
\mathcal{B}(D\to R\ell^+\nu_\ell,R\to P_1P_2)=\mathcal{B}(D\to R\ell^+\nu_\ell)\times\mathcal{B}(R\to P_1P_2), \label{Eq:BrD2RlvR2PP}
\end{eqnarray}
and this result is also a good approximation for wider resonances.   Above Eq. (\ref{Eq:BrD2RlvR2PP}) will be used in our analysis for the scalar resonant $D\to S(S\to P_1P_2)\ell^+\nu_\ell$  decays and the vector resonant  $D\to V(V\to P_1P_2)\ell^+\nu_\ell$ decays in Secs. \ref{Sec:S}  and \ref{Sec:V}, respectively. Relevant $\mathcal{B}(D\to R\ell^+\nu_\ell)$ and $\mathcal{B}(R\to P_1P_2)$ are also obtained by  the SU(3) flavor symmetry  in our later analysis.

\subsection{Meson multiplets}
Before giving the hadronic amplitudes based on the SU(3) flavor analysis,  we will  collect the representations for the multiplets of the  SU(3) flavor group first in this subsection.

Charmed mesons containing one heavy $c$ quark are flavor SU(3) antitriplets
\begin{eqnarray}
D_i=\Big( D^0(c\bar{u}),~D^+(c\bar{d}),~D^+_s(c\bar{s})\Big).
\end{eqnarray}
Light pseudoscalar meson ($P$) and vector meson ($V$)   octets and singlets under the SU(3) flavor symmetry of light $u,d,s$ quarks are \cite{He:2018joe}
\begin{eqnarray}
 P&=&\left(\begin{array}{ccc}
\frac{\pi^0}{\sqrt{2}}+\frac{\eta_8}{\sqrt{6}}+\frac{\eta_1}{\sqrt{3}} & \pi^+ & K^+ \\
\pi^- &-\frac{\pi^0}{\sqrt{2}}+\frac{\eta_8}{\sqrt{6}}+\frac{\eta_1}{\sqrt{3}}  & K^0 \\
K^- & \overline{K}^0 &-\frac{2\eta_8}{\sqrt{6}}+\frac{\eta_1}{\sqrt{3}}
\end{array}\right)\,,\\
V&=&\left(\begin{array}{ccc}
\frac{\rho^0}{\sqrt{2}}+\frac{\omega}{\sqrt{2}} & \rho^+ & K^{*+} \\
\rho^- &-\frac{\rho^0}{\sqrt{2}}+\frac{\omega}{\sqrt{2}} & K^{*0} \\
K^{*-} & \overline{K}^{*0} &\phi
\end{array}\right)\,,
\end{eqnarray}
where the $\eta$ and $\eta'$  are mixtures of $\eta_1=\frac{u\bar{u}+d\bar{d}+s\bar{s}}{\sqrt{3}}$ and $\eta_8=\frac{u\bar{u}+d\bar{d}-2s\bar{s}}{\sqrt{6}}$ with the mixing angle $\theta_P$
\begin{eqnarray}
\left(\begin{array}{c}
\eta\\
\eta'
\end{array}\right)\,
=
\left(\begin{array}{cc}
\mbox{cos}\theta_P&-\mbox{sin}\theta_P\\
\mbox{sin}\theta_P&\mbox{cos}\theta_P
\end{array}\right)\,\left(\begin{array}{c}
\eta_8\\
\eta_1
\end{array}\right)\,.
\end{eqnarray}\label{Eq:etamix}
And $\theta_P=[-20^\circ,-10^\circ]$ from  the Particle Data Group (PDG) \cite{PDG2022} will be used in our numerical  analysis.

The structures of the light scalar mesons  are   not fully understood
yet. Many suggestions are discussed, such as ordinary two-quark state, four-quark state, meson-meson bound state, molecular state,  glueball state, or hybrid state; for examples, see Refs. \cite{Dai:2018fmx,Maiani:2004uc,tHooft:2008rus,Pelaez:2003dy,Sun:2010nv,Oller:1997ti,Baru:2003qq,Cheng:2005nb,Achasov:1996ei}.
In this work, we will consider the two-quark and the four-quark scenarios for the scalar mesons below or near 1 GeV.
In the two-quark picture, the light scalar mesons can be written as   \cite{Momeni:2022gqb}
\begin{eqnarray}
S&=&\left(\begin{array}{ccc}
\frac{a^0_0}{\sqrt{2}}+\frac{\sigma}{\sqrt{2}} & a^+_0 & K^{+}_0 \\
a_0^- &-\frac{a_0^0}{\sqrt{2}}+\frac{\sigma}{\sqrt{2}} & K^{0}_0 \\
K^{-}_0 & \overline{K}^{0}_0 &f_0
\end{array}\right)\,.
\end{eqnarray}
The two isoscalars $f_0(980)$ and $f_0(500)$ are obtained by the mixing of $\sigma=\frac{u\bar{u}+d\bar{d}}{\sqrt{2}}$ and $f_0=s\bar{s}$,
\begin{eqnarray}
\left(\begin{array}{c}
f_0(980)\\
f_0(500)
\end{array}\right)\,
=
\left(\begin{array}{cc}
\mbox{cos}\theta_S&\mbox{sin}\theta_S\\
-\mbox{sin}\theta_S&\mbox{cos}\theta_S
\end{array}\right)\,\left(\begin{array}{c}
f_0\\
\sigma
\end{array}\right)\,,
\end{eqnarray}\label{Eq:f0mix2q}
where the three possible ranges of the mixing angle $\theta_S$ \cite{Cheng:2005nb,LHCb:2013dkk}, $25^\circ<\theta_S<40^\circ$, $140^\circ<\theta_S<165^\circ $ and $~-30^\circ<\theta_S<30^\circ $  will be analyzed in our numerical results.
In the four-quark picture, the light scalar mesons  are given as   \cite{Jaffe:1976ig,PDG2022}
\begin{eqnarray}
&&\sigma=u\bar{u}d\bar{d},~~~~~~~~~~~~~~~~~~~~~~~~~~f_0=(u\bar{u}+d\bar{d})s\bar{s}/\sqrt{2},\nonumber\\
&&a^0_0=(u\bar{u}-d\bar{d})s\bar{s}/\sqrt{2},~~~~~~~~~~~~~~~a^+_0=u\bar{d}s\bar{s},~~~~~~~~~~~~~~~~a^-_0=d\bar{u}s\bar{s},\nonumber\\
&& K^+_0=u\bar{s}d\bar{d},~~~~~~~~~K^0_0=d\bar{s}u\bar{u},~~~~~~~~~\overline{K}^0_0=s\bar{d}u\bar{u},~~~~~~~~~K^-_0=s\bar{u}d\bar{d},
\end{eqnarray}
and the two isoscalars are expressed as
\begin{eqnarray}
\left(\begin{array}{c}
f_0(980)\\
f_0(500)
\end{array}\right)\,
=
\left(\begin{array}{cc}
\mbox{cos}\phi_S&\mbox{sin}\phi_S\\
-\mbox{sin}\phi_S&\mbox{cos}\phi_S
\end{array}\right)\,\left(\begin{array}{c}
f_0\\
\sigma
\end{array}\right)\,,
\end{eqnarray}\label{Eq:f0mix4q}
where the constrained mixing angle $\phi_S=(174.6^{+3.4}_{-3.2})^\circ$ \cite{Maiani:2004uc}.

\subsection{ Nonresonant hadronic amplitudes } \label{sec:FFSU3orB}

Since the  hadronic amplitudes of the semileptonic  $D\to V/S\ell^+\nu_\ell$ decays based on the SU(3) flavor symmetry/breaking have been discussed in Ref. \cite{Wang:D2MlvSU3}, we will focus on the hadronic amplitudes of the nonresonant  $D\to P_1P_2\ell^+\nu_\ell$ decays in this subsection.

\begin{figure}[b]
\begin{center}
\includegraphics[scale=0.65]{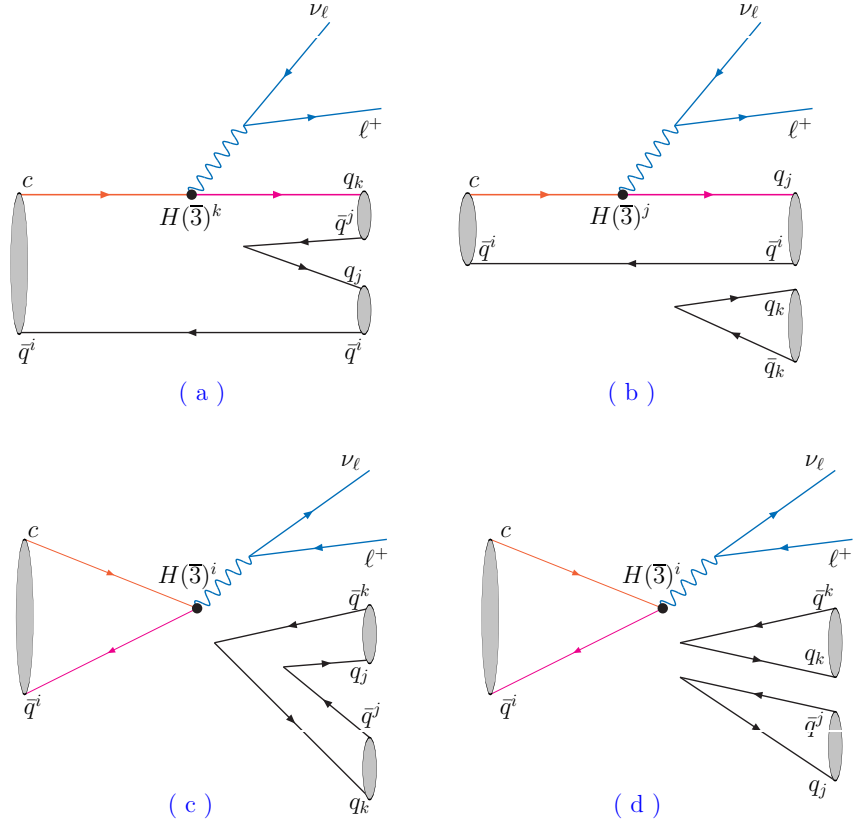}
\end{center}
\caption{Diagrams of the nonresonant  $D\to P_1P_2\ell^+\nu_\ell$ decays. }\label{fig1DG}
\end{figure}

In terms of  the SU(3) flavor symmetry, the quark current $\bar{q}_i\gamma^\mu(1-\gamma_5)c$ can be expressed as a SU(3) flavor  anti-triplet $(\bar{3})$,  and  the  effective Hamiltonian in Eq. (\ref{Heff}) is transformed as \cite{Geng:2017mxn}
\begin{eqnarray}
\mathcal{H}_{eff}(c\rightarrow q_i\ell^+\nu_\ell)&=&\frac{G_F}{\sqrt{2}}H(\bar{3})~\bar{\nu}_\ell\gamma_\mu(1-\gamma_5)\ell,\label{HeffSU3}
\end{eqnarray}
with $H(\bar{3})=(0,V_{cd},V_{cs})$.
The decay amplitude of the nonresonant $D\rightarrow P_1P_2\ell^+\nu_\ell$ decay can be written as
\begin{eqnarray}
\mathcal{A}(D\rightarrow P_1P_2\ell^+\nu_\ell)_N=\frac{G_F}{\sqrt{2}} H(D\to P_1P_2)_N~\bar{\nu}_\ell\gamma_\mu(1-\gamma_5)\ell,  \label{Eq:ASU3}
\end{eqnarray}
and the hadronic amplitude $H(D\to P_1P_2)_N$   can be
parameterized as
\begin{eqnarray}
H(D\to P_1P_2)_N =c_{10}D_iP^i_jP^j_kH(\bar{3})^k+c_{20}D_iP^i_jH(\bar{3})^jP^k_k +c_{30}D_iH(\bar{3})^iP^j_kP^k_j+c_{40}D_iH(\bar{3})^iP^k_kP^j_j,\label{Eq:AND2PPlv}
\end{eqnarray}
where $c_{i0}(i=1,2,3,4)$ are the nonperturbative coefficients under the SU(3) flavor symmetry. %
Feynman diagrams for the nonresonant $D\to P_1P_2\ell^+\nu_\ell$ decays
are displayed in Fig. \ref{fig1DG}.

 SU(3) flavor breaking effects  come from different masses of $u$, $d$ and $s$ quarks, and they
 will become useful once we have measurements of
several $D\to P_1P_2\ell^+\nu_\ell$ decays that are precise enough to see deviations from the SU(3) flavor symmetry.
The diagonalized mass matrix can be expressed as  \cite{Xu:2013dta,He:2014xha}
\begin{eqnarray}
 \left(\begin{array}{ccc}
m_u & 0&0\\
0 &m_d & 0 \\
0 & 0 &m_s
\end{array}\right)=\frac{1}{3}(m_u+m_d+m_s)I+\frac{1}{2}(m_u-m_d)X+\frac{1}{6}(m_u+m_d-2m_s)W,
\end{eqnarray}
with
\begin{eqnarray}
X= \left(\begin{array}{ccc}
1 & 0&0\\
0 &-1 & 0 \\
0 & 0 &0
\end{array}\right), ~~~~~~~~~~W= \left(\begin{array}{ccc}
1 & 0&0\\
0 &1 & 0 \\
0 & 0 &-2
\end{array}\right).
\end{eqnarray}
Compared with $s$ quark mass, the $u$ and $d$ quark masses are much smaller which can be ignored.
The SU(3) flavor breaking effects due to a nonzero $s$ quark mass dominate  the SU(3) breaking effects.  When $u$ and $d$ quark mass  difference is ignored, the residual SU(3) flavor symmetry becomes the isospin symmetry and the term proportional to $X$ can be dropped.  The identity $I$ part contributes to the $D\rightarrow P_1P_2\ell^+\nu_\ell$ decay amplitudes in a similar way
as that given in Eq. (\ref{Eq:ASU3}) which can be absorbed into the coefficients $c_{i0}$. Only the $W$
part will contribute to the SU(3) breaking effects. The SU(3) breaking contributions to the hadronic amplitudes due to the fact of $m_s\gg m_{u,d}$  are
\begin{eqnarray}
\Delta H(D\to P_1P_2)_N&=&c_{11}D_iW^i_aP^a_jP^j_kH(\bar{3})^k+c_{12}D_iP^i_jW^j_aP^a_kH(\bar{3})^k+c_{13}D_iP^i_jP^j_kW^k_aH(\bar{3})^a \nonumber\\
&+&c_{21}D_iW^i_aP^a_jH(\bar{3})^jP^k_k+c_{22}D_iP^i_jW^j_aH(\bar{3})^aP^k_k  \nonumber\\
&+&c_{31}D_iW^i_aH(\bar{3})^aP^j_kP^k_j +c_{32}D_iH(\bar{3})^iP^j_kW^k_aP^a_j\nonumber\\
&+&c_{41}D_iW^i_aH(\bar{3})^aP^k_kP^j_j,\label{Eq:AND2PPlvSU3B}
\end{eqnarray}
where $c_{ij}~(i,j=1,2,3,4)$ are the nonperturbative    SU(3) flavor  breaking coefficients.

Full hadronic amplitudes of the different nonresonant $D\to P_1P_2\ell^+\nu$ decays and their relations    under the SU(3) flavor symmetry/breaking  are
given in  Sec. \ref{sec:ND2PPlv}.

\section{Numerical results of the $D\to P_1P_2\ell^+\nu$ decays }
The branching ratios with the nonresonant contributions, the light scalar meson resonant contributions and the vector meson resonant contributions will  be  analyzed  in this section.
If not specially specified, the theoretical input parameters, such as the lifetimes and  the masses,   and  the experimental
data within the $2\sigma$ error bars from PDG \cite{PDG2022} will be used in our numerical analysis.

\subsection{ Nonresonant $D\to P_1P_2\ell^+\nu$ decays } \label{sec:ND2PPlv}

The  hadronic amplitudes  of the nonresonant $D\to P_1P_2\ell^+\nu_\ell$  decays  including both the SU(3) flavor symmetry and the SU(3) flavor  breaking terms are summarized in the second column of Tab. \ref{Tab:HD2PPlvAmp}, in which we can see the relations of different hadronic amplitudes. The following relations are held in both the SU(3) flavor symmetry and the SU(3) flavor breaking due to a strange quark mass:
\begin{eqnarray}
H(D^0\to \pi^-\overline{K}^0\ell^+\nu_\ell)_N&=&H(D^+\to \pi^+K^-\ell^+\nu_\ell)_N= \sqrt{2}H(D^0\to \pi^0K^-\ell^+\nu_\ell)_N=-\sqrt{2}H(D^+\to \pi^0\overline{K}^0\ell^+\nu_\ell)_N,  \nonumber\\
H(D^0\to \eta_8K^-\ell^+\nu_\ell)_N&=&H(D^+\to \eta_8\overline{K}^0\ell^+\nu_\ell)_N,\nonumber\\
H(D^0\to \eta_1K^-\ell^+\nu_\ell)_N&=&H(D^+\to \eta_1\overline{K}^0\ell^+\nu_\ell)_N, \nonumber\\
H(D^+_s\to K^+K^-\ell^+\nu_\ell)_N&=&H(D^+_s\to K^0\overline{K}^0\ell^+\nu_\ell)_N, \nonumber\\
H(D^0\to K^-K^0\ell^+\nu_\ell)_N&=&H(D^+\to \overline{K}^0K^0\ell^+\nu_\ell)_N-H(D^+\to K^+K^-\ell^+\nu_\ell)_N, \nonumber\\
H(D^+_s\to K^+\pi^-\ell^+\nu_\ell)_N&=&-\sqrt{2}H(D^+_s\to K^0\pi^0\ell^+\nu_\ell)_N. \label{Eq:R}
\end{eqnarray}
\begin{table}[ht]
\renewcommand\arraystretch{1.1}
\tabcolsep 0.05in
\centering
\caption{The hadronic amplitudes  for the $D\to P_1P_2\ell^+\nu_\ell$ decays. $C_1\equiv c_{10}+c_{11}+c_{12}-2c_{13}$, $C_2\equiv c_{20}+c_{21}-2c_{22}$,  $C_3\equiv c_{30}-2c_{31}$,  $C_4\equiv c_{40}-2c_{41}$, and
$[C^{',''}]_R$ denotes the contributions come from the  decays with $R$ resonances.  }\vspace{0.1cm}
{\footnotesize
\begin{tabular}{llcc}  \hline
~~~~~Decay modes~~~~~  & Nonresonant   hadronic amplitudes    & Scalar resonant ones & Vector resonant ones \\\hline
\multicolumn{4}{l}{{\color{blue}$c \to s\ell^+\nu_\ell$:}}\\\hline
$D^0\to \pi^-\overline{K}^0\ell^+\nu_\ell$          &$C_1$ &                                                  $\big[C'_1\big]_{K^-_0}$                                                   &$\big[C''_1\big]_{K^{*-}}$\\\hline
$D^0\to \pi^0K^-\ell^+\nu_\ell$                     &$\frac{1}{\sqrt{2}}C_1$                                 &$\big[\frac{1}{\sqrt{2}}C'_1\big]_{K^-_0}$                                 &$\big[\frac{1}{\sqrt{2}}C''_1\big]_{K^{*-}}$\\\hline
$D^0\to \eta_8K^-\ell^+\nu_\ell$                    &$-\frac{1}{\sqrt{6}}C_1$~$+\sqrt{6}c_{12}$                               &$\cdots$                                                 &$\cdots$\\\hline
$D^0\to \eta_1K^-\ell^+\nu_\ell$                    &$\frac{2}{\sqrt{3}}\big(C_1+\frac{3}{2}C_2\big)$~$-\sqrt{3}c_{12}$       &$\cdots$                                                 &$\cdots$ \\\hline
$D^+\to \pi^+K^-\ell^+\nu_\ell$                     &$C_1$                                                   &$\big[C'_1\big]_{\overline{K}^0_0}$                                                   &$\big[C''_1\big]_{K^{*0}}$\\\hline
$D^+\to \pi^0\overline{K}^0\ell^+\nu_\ell$          &$-\frac{1}{\sqrt{2}}C_1$                               &$\big[\frac{1}{\sqrt{2}}C'_1\big]_{\overline{K}^0_0}$                                &$\big[\frac{1}{\sqrt{2}}C''_1\big]_{K^{*0}}$\\\hline
$D^+\to \eta_8\overline{K}^0\ell^+\nu_\ell$         &$-\frac{1}{\sqrt{6}}C_1$~$+\sqrt{6}c_{12}$                               &$\cdots$                                                &$\cdots$\\\hline
$D^+\to \eta_1\overline{K}^0\ell^+\nu_\ell$         &$\frac{2}{\sqrt{3}}\big(C_1+\frac{3}{2}C_2\big)$~$-\sqrt{3}c_{12}$      &$\cdots$                                                &$\cdots$\\\hline
$D^+_s\to K^+K^-\ell^+\nu_\ell$                     &$C_1+2C_3$ ~$-3c_{11}-c_{32}$                                           &$\big[cos^2\theta_S~C'_1\big]_{f_0(980)}$               &$\big[C''_1\big]_{\phi}$\\\hline
$D^+_s\to K^0\overline{K}^0\ell^+\nu_\ell$          &$C_1+2C_3$ ~$-3c_{11}-c_{32}$                                            &$\big[cos^2\theta_S~C'_1\big]_{f_0(980)}$               &$\big[C''_1\big]_{\phi}$\\\hline
$D^+_s\to \pi^0\pi^0\ell^+\nu_\ell$                 &$\sqrt{2}C_3$~$+\sqrt{2}c_{32}$        &$^{\big[sin\theta_Scos\theta_SC'_1\big]_{f_0(980)}}_{\big[-sin\theta_Scos\theta_SC'_1\big]_{f_0(500)}}$     &$\cdots$\\\hline
$D^+_s\to \pi^+\pi^-\ell^+\nu_\ell$                 &$2(C_3+c_{32})$              &$^{\big[\sqrt{2}sin\theta_Scos\theta_SC'_1\big]_{f_0(980)}}_{\big[-\sqrt{2}sin\theta_Scos\theta_SC'_1\big]_{f_0(500)}}$      &$\cdots$\\\hline
$D^+_s\to \eta_8\eta_8\ell^+\nu_\ell$               &$\frac{2\sqrt{2}}{3}\big(C_1+\frac{3}{2}C_3\big)$~$-\sqrt{2}\big(2c_{11}+2c_{12}+c_{32}\big)$      &$\cdots$                                                &$\cdots$\\\hline
$D^+_s\to \eta_1\eta_1\ell^+\nu_\ell$               &$\frac{\sqrt{2}}{3}(C_1+3C_2+3C_3+9C_4)$ ~$-\sqrt{2}(c_{11}+c_{12}+3c_{21})$              &$\cdots$                                                &$\cdots$\\\hline
$D^+_s\to \eta_8\eta_1\ell^+\nu_\ell$               &$-\frac{2\sqrt{2}}{3}\big(C_1+\frac{3}{2}C_2\big)$~$+2\sqrt{2}(c_{11}+c_{12}+\frac{3}{2}c_{21}+c_{32})$     &$\cdots$                                                &$\cdots$\\\hline
%
%
%
%
%
%
\multicolumn{2}{l}{{\color{blue}$c \to d\ell^+\nu_\ell$:}}\\\hline
$D^0\to K^-K^0\ell^+\nu_\ell$                       &$C_1$ ~$-3(c_{12}-c_{13})$                                                 &$\big[C'_1\big]_{a_0(980)}$                              &$\cdots$\\\hline
$D^0\to \pi^0\pi^-\ell^+\nu_\ell$                   &$\cdots$                                              &$\cdots$                                                 &$\big[\frac{1}{\sqrt{2}}C''_1\big]_{\rho^-}$\\\hline
$D^0\to \eta_8\pi^-\ell^+\nu_\ell$                  &$\sqrt{\frac{2}{3}}C_1$~$+\sqrt{6}c_{13}$                                &\big[$\sqrt{\frac{2}{3}}C'_1\big]_{a_0(980)}$            &$\big[\frac{1}{\sqrt{6}}C''_1\big]_{\rho^-}$\\\hline
$D^0\to \eta_1\pi^-\ell^+\nu_\ell$                  &$\frac{2}{\sqrt{3}}\big(C_1+\frac{3}{2}C_2\big)$~$+\sqrt{3}(2c_{13}+3c_{22})$       &$\big[\frac{2}{\sqrt{3}}C'_1\big]_{a_0(980)}$            &$\big[\frac{1}{\sqrt{3}}C''_1\big]_{\rho^-}$\\\hline
$D^+\to \overline{K}^0K^0\ell^+\nu_\ell$            &$C_1+2C_3$~$-3(c_{12}-c_{13}-2c_{31})-c_{32}$         &$^{\big[\frac{1}{2}C'_1\big]_{a_0(980)}}_{\big[\frac{1}{\sqrt{2}}sin\theta_Scos\theta_SC'_1\big]_{f_0(980)}}$         &$\cdots$\\\hline
$D^+\to K^+K^-\ell^+\nu_\ell$                       &$2C_3$~$+6c_{31}-c_{32}$             &$^{\big[-\frac{1}{2}C'_1\big]_{a_0(980)}}_{\big[\frac{1}{\sqrt{2}}sin\theta_Scos\theta_SC'_1\big]_{f_0(980)}}$        &$\cdots$\\\hline
$D^+\to \pi^+\pi^-\ell^+\nu_\ell$                   &$C_1+2C_3$~$+3c_{13}+6c_{31}+2c_{32}$         &$^{\big[sin^2\theta_SC'_1\big]_{f_0(980)}}_{\big[cos^2\theta_SC'_1\big]_{f_0(500)}}$       &$\big[\frac{1}{2}C''_1\big]_{\rho^0,\omega}$\\\hline
$D^+\to \pi^0\pi^0\ell^+\nu_\ell$                   &$\frac{1}{\sqrt{2}}(C_1+2C_3)$~$+\frac{1}{\sqrt{2}}(3c_{13}+6c_{31}+2c_{32})$        &$^{\big[\frac{1}{\sqrt{2}}sin^2\theta_SC'_1\big]_{f_0(980)}}_{\big[\frac{1}{\sqrt{2}}cos^2\theta_SC'_1\big]_{f_0(500)}}$  &$\cdots$\\\hline
$D^+\to \eta_8\pi^0\ell^+\nu_\ell$                  &$-\frac{1}{\sqrt{3}}\big(C_1+C_2\big)$~$-\sqrt{3}\big(c_{13}+c_{22}\big)$                 &$\big[-\frac{1}{\sqrt{6}}C'_1\big]_{a_0(980)}$                                &$\cdots$\\\hline
$D^+\to \eta_1\pi^0\ell^+\nu_\ell$                  &$-\sqrt{\frac{2}{3}}\big(C_1+C_2\big)$~$-\frac{1}{\sqrt{6}}\big(6c_{13}+9c_{22}\big)$      &$\big[-\frac{1}{\sqrt{3}}C'_1\big]_{a_0(980)}$                                &$\cdots$\\\hline
$D^+\to \eta_8\eta_8\ell^+\nu_\ell$                 &$\frac{\sqrt{2}}{6}\big(C_1+6C_3\big)$ ~$+\frac{1}{\sqrt{2}}(c_{13}+6c_{31}-2c_{32})$                &$\cdots$                                                 &$\cdots$\\\hline
$D^+\to \eta_1\eta_1\ell^+\nu_\ell$                 &$\frac{\sqrt{2}}{3}(C_1+3C_2+3C_3+9C_4) +\sqrt{2}(c_{13}+3c_{22}+3c_{31}+9c_{41})$               &$\cdots$                                                 &$\cdots$\\\hline
$D^+\to \eta_8\eta_1\ell^+\nu_\ell$                 &$\frac{\sqrt{2}}{3}\big(C_1+\frac{3}{2}C_2\big)$~$+\sqrt{2}\big(c_{13}+\frac{3}{2}c_{22}+2c_{32}\big)$       &$\cdots$                                                 &$\cdots$\\\hline
$D^+_s\to K^+\pi^-\ell^+\nu_\ell$                   &$C_1$~$-3c_{11}+3c_{13}$                                                  &$\big[C'_1\big]_{K^0_0}$                                 &$\big[C''_1\big]_{K^{*0}}$\\\hline
$D^+_s\to K^0\pi^0\ell^+\nu_\ell$                   &$-\frac{1}{\sqrt{2}}C_1$~$-\frac{1}{\sqrt{2}}(-3c_{11}+3c_{13})$             &$\big[-\frac{1}{\sqrt{2}}C'_1\big]_{K^0_0}$                                &$\big[\frac{1}{\sqrt{2}}C''_1\big]_{K^{*0}}$\\\hline
$D^+_s\to \eta_8K^0\ell^+\nu_\ell$                  &$-\frac{1}{\sqrt{6}}C_1$~$+\frac{1}{\sqrt{6}}\big(3c_{11}+6c_{12}-3c_{13}\big)$                              &$\cdots$                                                 &$\cdots$\\\hline
$D^+_s\to \eta_1K^0\ell^+\nu_\ell$                  &$\frac{2}{\sqrt{3}}\big(C_1+\frac{3}{2}C_2\big)$~$-\sqrt{3}\big(2c_{11}+c_{12}-2c_{13}+3c_{21}-3c_{22}\big)$      &$\cdots$                                                 &$\cdots$\\\hline
\end{tabular}\label{Tab:HD2PPlvAmp}}
\end{table}
If assuming the SU(3) flavor  breaking effects are small and can be ignored,   more amplitude relations will be obtained.
Moreover, as shown in Fig. \ref{fig1DG},  the SU(3) flavor symmetry contributions of  Fig. \ref{fig1DG} (b-d) are suppressed by the Okubo-Zweig-Iizuka (OZI) rule \cite{Okubo:1963fa,Lipkin:1986bi,Lipkin:1996ny}.
If ignoring both the OZI suppressed   SU(3) flavor symmetry contributions and the SU(3) flavor breaking contributions, almost all hadronic amplitudes of the  nonresonant $D\to P_1P_2\ell^+\nu_\ell$  decays can be related by the coefficient $c_{10}$.

Since  the leptonic charged current  $\bar{\nu}_\ell\gamma_\mu(1-\gamma_5)\ell$ is the SU(3) flavor singlet, and it is  completely generic between different decay modes with certain $\ell=e$ or $\mu$. The same relations as the hadronic amplitudes listed in Tab. \ref{Tab:HD2PPlvAmp} are
valid in the decay amplitudes of the $D\to P_1P_2\ell^+\nu_\ell$ decays  and
the form factors of the $D\to P_1P_2$ transitions.
 For  the  nonresonant $D\to P_1P_2\ell^+\nu_\ell$ decays, only $\mathcal{B}(D^+\to \pi^+K^-\mu^+\nu_\mu)_N$ has been measured, and $\mathcal{B}(D^+\to \pi^+K^-e^+\nu_e)_N$ has been  upper limited. Because the  nonresonant $D\to P_1P_2\ell^+\nu_\ell$ decays have not been measured enough to reveal  the OZI suppressed SU(3) flavor symmetry contributions  and the  SU(3) symmetry breaking effects,  we ignore both of them in our analysis,  and then almost all hadronic amplitudes, form factors or decay amplitudes can be related by the SU(3) flavor symmetry  coefficient $c_{10}$.  The simple relations associated  by the  coefficient $c_{10}$ for $F_A$ given in Eq. (\ref{Eq:DB2PPlvdbr})    will be used  to obtain our numerical results.  Note that, for consistency, only the SU(3) flavor symmetry contributions will be considered in  the light scalar meson resonant $D\to S (S\to P_1P_2) \ell^+\nu_\ell$ decays and the vector meson resonant $D\to V (V\to P_1P_2) \ell^+\nu_\ell$ decays in Sec. \ref{Sec:S} and Sec. \ref{Sec:V}, respectively.

 The experimental data of $\mathcal{B}(D^+\to \pi^+K^-\mu^+\nu_\mu)_N$ within $2\sigma$ errors and the upper limit of $\mathcal{B}(D^+\to \pi^+K^-e^+\nu_e)_N$ at 90\% confidence level   from PDG \cite{PDG2022} are listed in the second column of Tab. \ref{Tab:BrD2MMlvcs}, which will be used to determine $c_{10}$ in the nonresonant $D^+\to \pi^+K^-\ell^+\nu_\ell$ decays,  and we obtain $|c_{01}|=12.95\pm3.75$ after considering  $2\sigma$ theoretical and experimental errors. Then many other branching ratios of the nonresonant $D\to P_1P_2\ell^+\nu_\ell$ decays can be predicted by using the constrained $c_{10}$ from the data of $\mathcal{B}(D^+\to \pi^+K^-\ell^+\nu_\ell)_N$  listed in the second column of Tab. \ref{Tab:BrD2MMlvcs}.
Our predictions  are listed in the third column of Tab. \ref{Tab:BrD2MMlvcs} for the $c \to s\ell^+\nu_\ell$ transitions and in the second column of  Tab. \ref{Tab:BrD2MMlvcd} for the $c \to d\ell^+\nu_\ell$ transitions.

From Tabs. \ref{Tab:BrD2MMlvcs}-\ref{Tab:BrD2MMlvcd}, one can see that many branching ratios of the nonresonant $D\to P_1P_2\ell^+\nu_\ell$ decays, such as  $\mathcal{B}(D^0\to \pi^-\overline{K}^0\ell^+\nu_\ell)_N$, $\mathcal{B}(D^0\to \pi^0K^-\ell^+\nu_\ell)_N$, $\mathcal{B}(D^+\to \pi^+K^-\ell^+\nu_\ell)_N$, $\mathcal{B}(D^+\to \pi^0\overline{K}^0\ell^+\nu_\ell)_N$, $\mathcal{B}(D^+_s\to K^+K^-\ell^+\nu_\ell)_N$, $\mathcal{B}(D^+_s\to K^0\overline{K}^0\ell^+\nu_\ell)_N$, $\mathcal{B}(D^+\to \pi^+\pi^-\ell^+\nu_\ell)_N$ and $\mathcal{B}(D^+\to \pi^0\pi^0\ell^+\nu_\ell)_N$,  are on the orders of $\mathcal{O}(10^{-3}-10^{-4})$, which could be measured by the BESIII,  LHCb, and BelleII experiments.  Nevertheless, for other decays, for example, the  nonresonant $D\to \eta P\ell^+\nu_\ell$ decays,  are strongly suppressed by the narrow phase spaces, the mixing angle $\theta_P$, or  the CKM matrix element $V_{cd}$,   their branching ratios are  on the orders of $\mathcal{O}(10^{-5}-10^{-7})$, and  many of them might be observed by the BESIII and Belle II experiments in the near future.

\begin{table}[h]
\renewcommand\arraystretch{1.5}
\tabcolsep 0.2in
\centering
\caption{ The experimental data and the SU(3) flavor symmetry  predictions
 of the nonresonant branching ratios  and the total branching ratios of the $D\to P_1P_2\ell^+\nu_\ell$ decays with the $c \to s\ell^+\nu_\ell$ transitions within the $2\sigma$ errors.  The experimental data are taken from PDG \cite{PDG2022},  `N' denotes the nonresonant contributions, and `T' denotes the total contributions including the non-resonance, the light scalar meson resonances as well as the vector meson resonances.  The same below. }\vspace{0.1cm}
{\footnotesize
\begin{tabular}{lcccc}  \hline
Branching ratios     &    Exp. data with N    &    Ones with N    &     Exp. data with T    &    Ones   with T \\\hline
%
%
%
$\mathcal{B}(D^0\to \pi^-\overline{K}^0e^+\nu_e)(\times10^{-2})$      &   $\cdots$        &   $0.076\pm0.041$     &   $1.44\pm0.08$             & $1.57\pm0.14$  \\
$\mathcal{B}(D^0\to \pi^0K^-e^+\nu_e)(\times10^{-2})$                 &   $\cdots$        &   $0.039\pm0.021$                  &   $1.6^{+2.6}_{-1.0}$       &   $0.80\pm0.07$\\
$\mathcal{B}(D^0\to \eta K^-e^+\nu_e)(\times10^{-6})$                 &   $\cdots$        &   $3.51\pm3.51$                    &    $\cdots$                         &  $3.51\pm3.51$\\
$\mathcal{B}(D^0\to \eta'K^-e^+\nu_e)(\times10^{-6})$                 &   $\cdots$        &   $4.03\pm2.17$                    &    $\cdots$                         &   $4.03\pm2.17$\\
$\mathcal{B}(D^+\to \pi^+K^-e^+\nu_e)(\times10^{-2})$                 & $<0.7$      &   $0.20\pm0.10$                    &   $4.02\pm0.36$             &   $4.06\pm0.30$\\
$\mathcal{B}(D^+\to \pi^0\overline{K}^0e^+\nu_e)(\times10^{-2})$      &   $\cdots$        &   $0.100\pm0.052$                  &    $\cdots$                         &   $2.01\pm0.15$\\
$\mathcal{B}(D^+\to \eta\overline{K}^0e^+\nu_e)(\times10^{-5})$       &   $\cdots$        &   $0.89\pm0.89$                    &    $\cdots$                         &   $0.89\pm0.89$\\
$\mathcal{B}(D^+\to \eta'\overline{K}^0e^+\nu_e)(\times10^{-5})$      &   $\cdots$        &   $1.03\pm0.55$                    &    $\cdots$                         &   $1.03\pm0.55$\\
$\mathcal{B}(D^+_s\to K^+K^-e^+\nu_e)(\times10^{-2})$                 &   $\cdots$        &   $0.034\pm0.018$                  &    $\cdots$                         &   $1.27\pm0.13$\\
$\mathcal{B}(D^+_s\to K^0\overline{K}^0e^+\nu_e)(\times10^{-3})$      &   $\cdots$        &   $0.33\pm0.18$                    &    $\cdots$                         &   $8.58\pm0.95$\\
$\mathcal{B}(D^+_s\to \pi^+\pi^-e^+\nu_e)(\times10^{-3})$             &   $\cdots$          &     $\cdots$                       &  $\cdots$                           &     $1.47\pm0.79$\\
$\mathcal{B}(D^+_s\to \pi^0\pi^0e^+\nu_e)(\times10^{-4})$             &   $\cdots$          &     $\cdots$                       &  $\cdots$                           &     $8.58\pm3.50$\\
$\mathcal{B}(D^+_s\to \eta \eta e^+\nu_e)(\times10^{-4})$             &   $\cdots$        &   $0.56\pm0.49$                    &    $\cdots$                         &   $0.56\pm0.49$\\
%
%
$\mathcal{B}(D^+_s\to \eta \eta'e^+\nu_e)(\times10^{-6})$           &   $\cdots$        &   $5.38\pm3.19$                    &  $\cdots$                           &   $5.38\pm3.19$\\\hline
%
%
%
%
%
$\mathcal{B}(D^0\to \pi^-\overline{K}^0\mu^+\nu_\mu)(\times10^{-2})$           &   $\cdots$                  &   $0.073\pm0.039$                        & $\cdots$                             & $1.47\pm0.13$   \\
$\mathcal{B}(D^0\to \pi^0K^-\mu^+\nu_\mu)(\times10^{-2})$                      &   $\cdots$                  &   $0.038\pm0.020$                        &  $\cdots$                            & $0.75\pm0.07$  \\
$\mathcal{B}(D^0\to \eta K^-\mu^+\nu_\mu)(\times10^{-6})$                      &   $\cdots$                  &   $3.18\pm3.18$                          &  $\cdots$                            & $3.18\pm3.18$   \\
$\mathcal{B}(D^0\to \eta'K^-\mu^+\nu_\mu)(\times10^{-6})$                      &   $\cdots$                  &   $2.76\pm1.49$                          &  $\cdots$                            &$2.76\pm1.49$    \\
$\mathcal{B}(D^+\to \pi^+K^-\mu^+\nu_\mu)(\times10^{-2})$                      &  $0.19\pm0.10$        &   $0.19\pm0.10$                          &   $3.65\pm0.68$              & $3.80\pm0.27$\\
$\mathcal{B}(D^+\to \pi^0\overline{K}^0\mu^+\nu_\mu)(\times10^{-2})$           &   $\cdots$                  &   $0.095\pm0.050$                        &   $\cdots$                           & $1.89\pm0.13$   \\
$\mathcal{B}(D^+\to \eta\overline{K}^0\mu^+\nu_\mu)(\times10^{-5})$            &   $\cdots$                  &   $0.81\pm0.81$                          &  $\cdots$                            &$0.81\pm0.81$    \\
$\mathcal{B}(D^+\to \eta'\overline{K}^0\mu^+\nu_\mu)(\times10^{-5})$           &   $\cdots$                  &   $0.71\pm0.38$                          &  $\cdots$                            &$0.71\pm0.38$    \\
$\mathcal{B}(D^+_s\to K^+K^-\mu^+\nu_\mu)(\times10^{-2})$                      &   $\cdots$                  &   $0.032\pm0.017$                        &   $\cdots$                           & $1.19\pm0.12$   \\
$\mathcal{B}(D^+_s\to K^0\overline{K}^0\mu^+\nu_\mu)(\times10^{-3})$           &   $\cdots$                  &   $0.30\pm0.16$                          &   $\cdots$                           & $8.02\pm0.88$   \\
$\mathcal{B}(D^+_s\to \pi^+\pi^-\mu^+\nu_\mu)(\times10^{-3})$                  &   $\cdots$                    &   $\cdots$                               &  $\cdots$                            & $1.25\pm0.69$ \\
$\mathcal{B}(D^+_s\to \pi^0\pi^0\mu^+\nu_\mu)(\times10^{-4})$                  &   $\cdots$                    &    $\cdots$                              &  $\cdots$                            & $7.34\pm3.09$    \\
$\mathcal{B}(D^+_s\to \eta \eta \mu^+\nu_\mu)(\times10^{-4})$                  &   $\cdots$                  &   $0.51\pm0.45$                          &     $\cdots$                         & $0.51\pm0.45$   \\
%
%
$\mathcal{B}(D^+_s\to \eta \eta'\mu^+\nu_\mu)(\times10^{-6})$                  &   $\cdots$                  &   $3.98\pm2.36$                          &     $\cdots$                         &$3.98\pm2.36$    \\\hline
\end{tabular}\label{Tab:BrD2MMlvcs}}
\end{table}

\begin{table}[h]
\renewcommand\arraystretch{1.3}
\tabcolsep 0.3in
\centering
\caption{ The experimental data and the SU(3) flavor symmetry  predictions of  the nonresonant branching ratios  and the total branching ratios of the  $D\to P_1P_2\ell^+\nu_\ell$ decays  with the $c \to d\ell^+\nu_\ell$ transitions within the $2\sigma$ errors. }\vspace{0.1cm}
{\footnotesize
\begin{tabular}{lccc}  \hline
Branching ratios         &      Ones with N      &       Exp. data with T      &      Ones with T \\\hline
%
%
%
$\mathcal{B}(D^0\to K^-K^0e^+\nu_e)(\times10^{-5})$                     &     $0.83\pm0.45$               &     $\cdots$                          &   $1.25\pm0.64$   \\
$\mathcal{B}(D^0\to \pi^0\pi^-e^+\nu_e)(\times10^{-3})$                 &     $0$                         &     $1.45\pm0.14$                &   $1.85\pm0.11$    \\
$\mathcal{B}(D^0\to \eta\pi^-e^+\nu_e)(\times10^{-5})$                  &     $4.34\pm2.68$               &     $\cdots$                           &   $16.38\pm5.10$   \\
$\mathcal{B}(D^0\to \eta'\pi^-e^+\nu_e)(\times10^{-5})$                 &     $0.39\pm0.26$               &     $\cdots$                              &   $0.57\pm0.35$  \\
$\mathcal{B}(D^+\to \overline{K}^0K^0e^+\nu_e)(\times10^{-5})$    &     $2.11\pm1.13$               &  $\cdots$                                 &     $3.31\pm1.69$\\
$\mathcal{B}(D^+\to K^+K^-e^+\nu_e)(\times10^{-5})$                     &     $\cdots$                    &   $\cdots$                         &     $1.31\pm0.63 $\\
$\mathcal{B}(D^+\to \pi^+\pi^-e^+\nu_e)(\times10^{-3})$           &     $0.26\pm0.14$               &     $2.45\pm0.20$                &     $3.08\pm0.51$\\
$\mathcal{B}(D^+\to \pi^0\pi^0e^+\nu_e)(\times10^{-4})$           &     $1.33\pm0.71$               &   $\cdots$                                &     $2.88\pm1.75$\\
$\mathcal{B}(D^+\to \eta\pi^0 e^+\nu_e)(\times10^{-5})$                 &     $5.68\pm3.50$               &     $\cdots$                              &     $9.68\pm4.49$\\
$\mathcal{B}(D^+\to \eta'\pi^0e^+\nu_e)(\times10^{-6})$           &     $5.21\pm3.46$               &    $\cdots$                               &     $8.28\pm5.00$\\
$\mathcal{B}(D^+\to \eta \eta e^+\nu_e)(\times10^{-6})$           &     $3.16\pm2.26$               &  $\cdots$                                 &     $3.16\pm2.26$\\
$\mathcal{B}(D^+\to \eta \eta'e^+\nu_e)(\times10^{-8})$        &     $3.96\pm2.37$               &   $\cdots$                                &     $3.96\pm2.37$\\
$\mathcal{B}(D^+_s\to K^+\pi^-e^+\nu_e)(\times10^{-3})$          &     $0.075\pm0.041$             &   $\cdots$                                &     $1.66\pm0.17$\\
$\mathcal{B}(D^+_s\to K^0\pi^0e^+\nu_e)(\times10^{-4})$          &     $0.38\pm0.21$               &  $\cdots$                                 &     $8.24\pm0.85$\\
$\mathcal{B}(D^+_s\to \eta K^0 e^+\nu_e)(\times10^{-5})$       &     $1.70\pm1.06$               &   $\cdots$                                &     $1.70\pm1.06$\\
$\mathcal{B}(D^+_s\to \eta'K^0e^+\nu_e)(\times10^{-7})$              &     $5.21\pm3.47$               &     $\cdots$                           &     $5.21\pm3.47$\\\hline
%
%
%
%
$\mathcal{B}(D^0\to K^-K^0\mu^+\nu_\mu)(\times10^{-5})$               &     $0.76\pm0.43$               &    $\cdots$                          &    $1.11\pm0.57$    \\
$\mathcal{B}(D^0\to \pi^0\pi^-\mu^+\nu_\mu)(\times10^{-3})$           &     $0$                         &     $\cdots$                           &   $1.76\pm0.10$    \\
$\mathcal{B}(D^0\to \eta\pi^-\mu^+\nu_\mu)(\times10^{-5})$            &     $4.13\pm2.55$               &     $\cdots$                           &   $15.04\pm4.76$   \\
$\mathcal{B}(D^0\to \eta'\pi^-\mu^+\nu_\mu)(\times10^{-5})$           &     $0.34\pm0.23$               &     $\cdots$                           &    $0.50\pm0.31$   \\
$\mathcal{B}(D^+\to \overline{K}^0K^0\mu^+\nu_\mu)(\times10^{-5})$   &     $1.93\pm1.04$               &  $\cdots$                                 &     $2.94\pm1.50$\\
$\mathcal{B}(D^+\to K^+K^-\mu^+\nu_\mu)(\times10^{-5})$          &     $\cdots$                    &    $\cdots$                          &     $1.09\pm0.53 $\\
$\mathcal{B}(D^+\to \pi^+\pi^-\mu^+\nu_\mu)(\times10^{-3})$      &     $0.25\pm0.14$               &     $\cdots$                           &     $2.92\pm0.48$\\
$\mathcal{B}(D^+\to \pi^0\pi^0\mu^+\nu_\mu)(\times10^{-4})$      &     $1.29\pm0.69$               &     $\cdots$                           &     $2.68\pm1.65$\\
$\mathcal{B}(D^+\to \eta\pi^0 \mu^+\nu_\mu)(\times10^{-5})$      &     $5.40\pm3.33$               &     $\cdots$                           &     $8.71\pm4.16$\\
$\mathcal{B}(D^+\to \eta'\pi^0\mu^+\nu_\mu)(\times10^{-6})$      &     $4.67\pm3.10$               &     $\cdots$                           &     $7.23\pm4.37$\\
$\mathcal{B}(D^+\to \eta \eta \mu^+\nu_\mu)(\times10^{-6})$      &     $2.83\pm2.02$               &     $\cdots$                           &     $2.83\pm2.02$\\
%
%
$\mathcal{B}(D^+\to \eta \eta'\mu^+\nu_\mu)(\times10^{-8})$          &     $2.43\pm1.46$               &     $\cdots$                           &     $2.43\pm1.46$\\
$\mathcal{B}(D^+_s\to K^+\pi^-\mu^+\nu_\mu)(\times10^{-3})$          &     $0.072\pm0.039$             &     $\cdots$                           &     $1.58\pm0.16$\\
$\mathcal{B}(D^+_s\to K^0\pi^0\mu^+\nu_\mu)(\times10^{-4})$          &     $0.36\pm0.20$               &     $\cdots$                           &     $7.81\pm0.80$\\
$\mathcal{B}(D^+_s\to \eta K^0 \mu^+\nu_\mu)(\times10^{-5})$         &     $1.57\pm0.98$               &     $\cdots$                           &     $1.57\pm0.98$\\
$\mathcal{B}(D^+_s\to \eta' K^0\mu^+\nu_\mu)(\times10^{-7})$         &     $4.08\pm2.72$               &     $\cdots$                           &     $4.08\pm2.72$\\\hline
\end{tabular}\label{Tab:BrD2MMlvcd}}
\end{table}

\clearpage
\subsection{$D\to S (S\to P_1P_2) \ell^+\nu_\ell$ decays}\label{Sec:S}
We will analyze the  $D\to P_1P_2\ell^+\nu_\ell$  decays with the light scalar resonances in this subsection.  As given in Eq. (\ref{Eq:BrD2RlvR2PP}), their branching ratios can be obtained by using $\mathcal{B}(D\to S\ell^+\nu_\ell)$ and $\mathcal{B}(S\to P_1P_2)$.
The detailed analysis of $\mathcal{B}(D\to S\ell^+\nu_\ell)$ by the SU(3) flavor symmetry can be found in Ref. \cite{Wang:D2MlvSU3}.

\subsubsection{Branching ratios of the $S\to P_1P_2$ decays}

As for the $S\to P_1P_2$ decays, the partial decay widths  can be written as \cite{Cheng:2020ipp}
\begin{eqnarray}
\Gamma(S\to P_1P_2)=\frac{p_c}{8\pi m^2_S}g^2_{S\to P_1P_2},
\end{eqnarray}
where the center-of-mass  momentum $p_c\equiv\frac{\sqrt{\lambda(m_S^2,m_{P_1}^2,m_{P_2}^2)}}{2m_S}$, and  $g_{S\to P_1P_2}$ is the strong coupling constant.
With the SU(3) flavor symmetry, the strong coupling constant can be parametrized as
 \begin{eqnarray}
g^{2q}_{S\to P_1P_2}=g_2S^i_jP^k_iP^j_k \label{Eq:gS2PP2q}
\end{eqnarray}
for the two-quark scalar states, and
\begin{eqnarray}
g^{4q}_{S\to P_1P_2}=g_4S^{im}_{jn}P^j_iP^n_m+g'_4S^{im}_{jm}P^n_iP^j_n \label{Eq:gS2PP4q}
\end{eqnarray}
for the four-quark scalar states, where $g_2$, $g_4$  and $g'_4$ are the nonperturbative parameters.  The strong coupling constants of these decays
are listed in the second and third columns of  Tab. \ref{Tab:S2PPR} for  the two-quark scalar states and the four-quark scalar states, respectively.
\begin{table}[htb]
\renewcommand\arraystretch{1.1}
\tabcolsep 0.25in
\centering
\caption{The strong coupling constants of the $S\to P_1P_2$ decays  by the SU(3) flavor symmetry. }\vspace{0.5cm}
\begin{tabular}{lccc}  \hline
Strong couplings  & Ones for two-quark state & Ones for four-quark state \\\hline
$g_{K_0^-\to \pi^0K^-}$     &      $\frac{1}{\sqrt{2}}~g_2$      &      $-\frac{1}{\sqrt{2}}g_4$\\
$g_{K_0^-\to \pi^-\overline{K}^0}$     &      $g_2$      &      $g_4$\\\hline
$g_{\overline{K}_0^0\to \pi^+K^-}$     &      $g_2$      &      $g_4$ \\
$g_{\overline{K}_0^0\to \pi^0\overline{K}^0}$     &      $-\frac{1}{\sqrt{2}}~g_2$      &      $\frac{1}{\sqrt{2}}g_4$\\\hline
$g_{a_0(980)^-\to \eta\pi^-}$     &      $2~g_2~\big(\frac{1}{\sqrt{6}}cos\theta_P-\frac{1}{\sqrt{3}}sin\theta_P\big)$      &      $2~g'_4~\big(\frac{1}{\sqrt{6}}cos\theta_P-\frac{1}{\sqrt{3}}sin\theta_P\big)$ \\
$g_{a_0(980)^-\to \eta'\pi^-}$      &      $2~g_2~\big(\frac{1}{\sqrt{6}}sin\theta_P+\frac{1}{\sqrt{3}}cos\theta_P\big)$      &      $2~g'_4~\big(\frac{1}{\sqrt{6}}sin\theta_P+\frac{1}{\sqrt{3}}cos\theta_P\big)$\\
$g_{a_0(980)^-\to K^0K^-}$     &      $g_2$      &      $g_4$\\\hline
$g_{a_0(980)^0\to \eta\pi^0}$     &      $g_2~\big(\frac{1}{\sqrt{3}}cos\theta_P-\sqrt{\frac{2}{3}}sin\theta_P\big)$      &      $g'_4~\big(\frac{1}{\sqrt{6}}cos\theta_P-\frac{1}{\sqrt{3}}sin\theta_P\big)$ \\
$g_{a_0(980)^0\to \eta'\pi^0}$      &      $g_2~\big(\frac{1}{\sqrt{3}}sin\theta_P+\sqrt{\frac{2}{3}}cos\theta_P\big)$      &      $g'_4~\big(\frac{1}{\sqrt{6}}sin\theta_P+\frac{1}{\sqrt{3}}cos\theta_P\big)$\\
$g_{a_0(980)^0\to K^+K^-}$      &      $\frac{1}{\sqrt{2}}~g_2$      &      $\frac{1}{\sqrt{2}}~g_4$\\
$g_{a_0(980)^0\to K^0\overline{K}^0}$      &      $-\frac{1}{\sqrt{2}}~g_2$      &      $-\frac{1}{\sqrt{2}}~g_4$\\\hline
$g_{f_0(980)\to \pi^+\pi^-}$      &      $\sqrt{2}~g_2~sin\theta_S$   &      $\sqrt{2}~g'_4~cos\phi_S+g_4sin\phi_S$\\
$g_{f_0(980)\to \pi^0\pi^0}$      &      $g_2~sin\theta_S$            &      $g'_4~cos\phi_S-\frac{1}{\sqrt{2}}g_4sin\phi_S$\\
$g_{f_0(980)\to K^+K^-}$          &      $g_2~cos\theta_S$            &      $\frac{1}{\sqrt{2}}g_4cos\phi_S$ \\
$g_{f_0(980)\to K^0\overline{K}^0}$    &      $g_2~cos\theta_S$            &      $\frac{1}{\sqrt{2}}g_4cos\phi_S$\\
\hline
$g_{f_0(500)\to \pi^+\pi^-}$      & $\sqrt{2}~g_2~cos\theta_S$        &      $-\sqrt{2}~g'_4~sin\phi_S+g_4cos\phi_S$\\
$g_{f_0(500)\to \pi^0\pi^0}$      &      $g_2~cos\theta_S$            &      $-g'_4~sin\phi_S-\frac{1}{\sqrt{2}}g_4cos\phi_S$\\\hline
\end{tabular}\label{Tab:S2PPR}
\end{table}

Since the width determination is very model dependent, there are not accurate values about the decay widths of $a_0(980)$, $f_0(980)$ and $f_0(500)$ mesons in Ref. \cite{PDG2022}. Therefore, it is difficult to  obtain accurate $\mathcal{B}(S\to P_1P_2)$ in terms of  $\Gamma(S\to P_1P_2)/\Gamma_S$, where $\Gamma_S$ is the decay width of scalar meson.
We assume the light scalar mesons decay dominantly into pairs of pseudoscalar mesons and all other decay channels are negligible,   and  then one can obtain $\mathcal{B}(S\to P_1P_2)$ without the  decay width values of the light scalar mesons, for example, $\mathcal{B}(f_0(500)\to \pi^+\pi^-)\approx\frac{\Gamma(f_0(500)\to \pi^+\pi^-)}{\Gamma(f_0(500)\to \pi^+\pi^-)+\Gamma(f_0(500)\to \pi^0\pi^0)}$.

In the two-quark picture, the parameter $g_2$ is cancelled in the branching ratios. Therefore,  $\mathcal{B}(K_0 \to \pi K ,a_0(980)\to KK,f_0(500)\to \pi\pi)$   only  depend on the masses of relevant mesons,  $\mathcal{B}(a_0(980)\to \eta'\pi,\eta'\pi)$   depend on the meson masses and the mixing angle  $\theta_P$, and  $\mathcal{B}(f_0(980)\to \pi\pi,KK)$   depend on the meson masses  and the mixing angle $\theta_S$.   The numerical results of $\mathcal{B}(S\to P_1P_2)$ in the two-quark picture are listed in the second column of Tab. \ref{Tab:BrS2PP}. One can see that the branching ratios of the $K_0,a_0(980),f_0(500)$ decays are accurately predicted;  nevertheless, $\mathcal{B}(f_0(980)\to \pi\pi,KK)$ are predicted with large error due to the indeterminate mixing angle $\theta_S$.    The three possible ranges for the mixing angle $\theta_S$,  $25^\circ<\theta_S<40^\circ$, $140^\circ<\theta_S<165^\circ$ and $~-30^\circ<\theta_S<30^\circ $ \cite{Cheng:2005nb,LHCb:2013dkk},  have been considered, and the predictions of $\mathcal{B}(f_0(980)\to \pi\pi,KK)$ are quite dependent on the mixing angle $\theta_S$.

In the third column of Tab. \ref{Tab:BrS2PP}, we also give the predictions with two-quark picture  of $\mathcal{B}(S\to P_1P_2)$ further constrained from the relevant experimental data of $\mathcal{B}(D\to S\ell^+\nu_\ell,S\to P_1P_2)$ listed in later Tabs. \ref{Tab:D2SlvS2PPcs}-\ref{Tab:D2SlvS2PPcd}.  The predictions of $\mathcal{B}(f_0(980) \to P_1P_2)$ are quite accurate when $\theta_S$ is further constrained from $[25^\circ,40^\circ]$ to $[25^\circ,36^\circ]$,  from $[140^\circ,165^\circ]$ to $[144^\circ,151^\circ]$ and from $|\phi_S|\leq30^\circ$  to  $22^\circ\leq|\phi_S|\leq30^\circ$  by the relevant experimental data of  $\mathcal{B}(D\to S\ell^+\nu_\ell,S\to P_1P_2)$  with $2\sigma$ errors.   Since $\theta_S$ in the two-quark picture has been further constrained by $\mathcal{B}(D\to S\ell^+\nu_\ell,S\to P_1P_2)$, the predictions of $\mathcal{B}(f(980)\to \pi\pi,KK)$ are more accurate as listed in the third column of Tab. \ref{Tab:BrS2PP}. Other  $\mathcal{B}(S\to P_1P_2)$  are not further constrained from the  data of $\mathcal{B}(D\to S\ell^+\nu_\ell,S\to P_1P_2)$, so we do not list them in the third column of Tab. \ref{Tab:BrS2PP}.

In the four-quark picture, the two nonperturbative parameters  $g_4$ and $g'_4$  in the $a_0(980),f_0(980), f_0(500)$ decays, and $|g'_4/g_4|=0.61\pm0.13$ are obtained by the data $\Gamma(a_0(980)\to K\bar{K})/\Gamma(a_0(980)\to \eta\pi)=0.177\pm0.048$ from PDG \cite{PDG2022}.  In this work, we treat $g_4$ and $g'_4$ as real number,  then  two possible cases ($g'_4/g_4>0$ and $g'_4/g_4<0$) are analyzed.  The numerical results  with the four-quark picture are listed in the last column of Tab. \ref{Tab:BrS2PP}.  As for   $\mathcal{B}(f_0(980)\to \pi\pi)$ and $\mathcal{B}(f_0(500)\to \pi\pi)$,  very large errors  come from the mixing angles $\phi_S$,  and they are obviously different in the $g'_4/g_4>0$ and $g'_4/g_4<0$ cases.   In general, there is a relative strong phase between $g'_4$ and $g_4$;  therefore, the common relevant branching ratios are between those in the $g'_4/g_4>0$ case and those in the $g'_4/g_4<0$ case.  In addition, $\mathcal{B}(K_0\to P_1P_2)$  are the same in both the two-quark and four-quark pictures.

\subsubsection{Branching ratios of the $D\to S(S\to P_1P_2)\ell^+\nu_\ell$ decays }

Then $\mathcal{B}(D\to S\ell^+\nu_\ell,S\to P_1P_2)$ can be obtained in terms of $\mathcal{B}(S\to P_1P_2)$ listed in Tab. \ref{Tab:BrS2PP} and the expressions of $\mathcal{B}(D\to S\ell^+\nu_\ell)$ given in Ref. \cite{Wang:D2MlvSU3}.  Using the experimental data of  $\mathcal{B}(D^+_s\to f_0(980)e^+\nu_e)=(2.3\pm0.8)\times10^{-3}$ \cite{PDG2022} as well as $\mathcal{B}(D\to S\ell^+\nu_\ell,S\to P_1P_2)$ listed in the second columns of Tabs.  \ref{Tab:D2SlvS2PPcs} and \ref{Tab:D2SlvS2PPcd}.  The numerical results of $\mathcal{B}(D\to S\ell^+\nu_\ell,S\to P_1P_2)$ with $2\sigma$ errors for the two-quark  and four-quark pictures are given in Tab. \ref{Tab:D2SlvS2PPcs} and  \ref{Tab:D2SlvS2PPcd} for the  $c \to s\ell^+\nu_\ell$ and $c \to d\ell^+\nu_\ell$  transitions, respectively.
Our comments on the results are as follows.
\begin{itemize}
\item The experimental lower limits of $\mathcal{B}(D^0\to a_0(980)^-e^+\nu_e,~a_0(980)^-\to \eta\pi^-)$ and $\mathcal{B}(D^+\to f_0(500)e^+\nu_e,~f_0(500)\to \pi^+\pi^-)$ have not been used to constrain  the predictions of  $\mathcal{B}(D\to S\ell^+\nu_\ell,S\to P_1P_2)$,  since the two lower limits of the  SU(3) flavor symmetry predictions are slightly lower than their experimental data in both the two-quark  and four-quark pictures. For $\mathcal{B}(D^0\to a_0(980)^-e^+\nu_e,~a_0(980)^-\to \eta\pi^-)$,  one can see that the prediction in the two-quark picture agrees with experimental data within $2\sigma$ error bars;  nevertheless, the prediction in the four-quark picture is smaller, which only agrees with experimental data within $3\sigma$ error bars.  As for $\mathcal{B}(D^+\to f_0(500)e^+\nu_e,~f_0(500)\to \pi^+\pi^-)$, the prediction   in the two-quark picture is much smaller than its experimental lower limit with $2\sigma$ error,  nevertheless,  the prediction with $\frac{g'_4}{g_4}>0$  ($\frac{g'_4}{g_4}<0$ ) in the four-quark picture  agrees with its data within $2\sigma$ ($3\sigma$) error bars.  Therefore, in the later analysis of total contributions to  $\mathcal{B}(D\to P_1P_2\ell^+\nu_\ell )$, the predictions of $\mathcal{B}(D\to S\ell^+\nu_\ell,S\to P_1P_2)$  with $\frac{g'_4}{g_4}>0$  in the four-quark picture will be used.

\item  In the two-quark picture, though  the mixing angle $\theta_S$ only appears in the $D\to P_1P_2\ell^+\nu_\ell $ decays with $f_0(980)$ and  $f_0(500)$  resonances, all other predictions of the branching ratios are slightly affected by the experimental constraints. So we list all predictions in the three possible ranges of  the mixing angle $\theta_S$ in the third through fifth columns  of Tabs. \ref{Tab:D2SlvS2PPcs} and \ref{Tab:D2SlvS2PPcd}. One can see that all of the predictions that included the decays with $f_0(980)$ and  $f_0(500)$  resonances are similar in the three possible ranges of the mixing angle $\theta_S$.  As mentioned before, $\theta_S$ is constrained from $[25^\circ,40^\circ]$ to $[25^\circ,36^\circ]$,  from $[140^\circ,165^\circ]$ to $[144^\circ,151^\circ]$ and from $|\phi_S|\leq30^\circ$  to  $22^\circ\leq|\phi_S|\leq30^\circ$  by the relevant experimental data with $2\sigma$ errors.

\item A lot of the branching ratio predictions are quite different between the two-quark picture and the four-quark picture.  Present datum of $\mathcal{B}(D^+\to f_0(500)e^+\nu_e,~f_0(500)\to \pi^+\pi^-)$ favors  the four-quark picture of scalar mesons.  $\mathcal{B}(D\to S\ell^+\nu_\ell,S\to P_1P_2)$ with the $c \to s\ell^+\nu_\ell$ transitions are predicted on the order of $\mathcal{O}(10^{-3}-10^{-4})$. Due to the CKM  matrix element $V_{cd}$ suppressed,   $\mathcal{B}(D\to S\ell^+\nu_\ell,S\to P_1P_2)$ with the $c \to d\ell^+\nu_\ell$ transitions are predicted on the order of $\mathcal{O}(10^{-4}-10^{-6})$.

\item Some branching ratios of the  $D\to S (S\to P_1P_2) \ell^+\nu_\ell$ decays have been obtained in Refs. \cite{Shi:2017pgh,Shi:2021bvy}.
$\mathcal{B}(D^+\to Se^+\nu_e, S\to \pi^+\pi^-)= (6.99\pm2.46)\times 10^{-4}$ \cite{Shi:2017pgh}, $\mathcal{B}(D^+\to S\mu^+\nu_\mu, S\to \pi^+\pi^-)= (7.20\pm2.52)\times 10^{-4}$ \cite{Shi:2017pgh}, and  $\mathcal{B}(D^0\to a_0(980)^-\ell^+\nu_\ell, a_0(980)^-\to \eta\pi^-)= (1.36\pm0.21)\times 10^{-4}$ \cite{Shi:2021bvy}.
Our predictions in the four-quark picture of $\mathcal{B}(D^+\to S\ell^+\nu_\ell, S\to \pi^+\pi^-)$ are consistent with ones in Ref. \cite{Shi:2017pgh};  our predictions in the two-quark picture of $\mathcal{B}(D^0\to a_0(980)^-\ell^+\nu_\ell, a_0(980)^-\to \eta\pi^-)$ are consistent with ones in Ref. \cite{Shi:2021bvy}; nevertheless,  our predictions in  the four-quark picture are smaller than  ones in Ref. \cite{Shi:2021bvy}.

\end{itemize}

\begin{table}[b]
\renewcommand\arraystretch{1.35}
\tabcolsep 0.15in
\centering
\caption{Branching ratios of the $S\to P_1P_2$ decays within $2\sigma$ errors. The results   are obtained by the SU(3) flavor symmetry relations and  $\Gamma(a_0(980)\to K\bar{K})/\Gamma(a_0(980)\to \eta\pi)=0.177\pm0.048$ \cite{PDG2022}.   $^\dag$denotes the results  with $\frac{g'_4}{g_4}>0$, and $^\sharp$denotes ones with $\frac{g'_4}{g_4}<0$.}\vspace{0.5cm}
\begin{tabular}{llll}  \hline
Branching ratios  & ones with $2q$ state in $S_1$ case & ones with $2q$ state in $S_2$ case &ones with   $4q$ state\\
$\mathcal{B}(K_0^-\to \pi^0K^-)$            &       $0.34\pm0.00$      &&         $0.34\pm0.00$     \\
$\mathcal{B}(K_0^-\to \pi^-\overline{K}^0)$            &       $0.66\pm0.00$      &&         $0.66\pm0.00$     \\\hline
$\mathcal{B}(\overline{K}_0^0\to \pi^+K^-)$            &       $0.67\pm0.00$      & &         $0.67\pm0.00$     \\
$\mathcal{B}(\overline{K}_0^0\to \pi^0\overline{K}^0)$            &       $0.33\pm0.00$      &&          $0.33\pm0.00$     \\\hline
$\mathcal{B}(a_0(980)^-\to \eta\pi^-)$       &       $0.64\pm0.04$     &&       $0.86\pm0.03$      \\
$\mathcal{B}(a_0(980)^-\to \eta'\pi^-)$      &       $0.03\pm0.01$     & &      $0.04\pm0.01$      \\
$\mathcal{B}(a_0(980)^-\to K^0K^-)$      &       $0.33\pm0.03$         &&       $0.10\pm0.02$      \\\hline
$\mathcal{B}(a_0(980)^0\to \eta\pi^0)$      &       $0.60\pm0.04$      & &      $0.67\pm0.06$       \\
$\mathcal{B}(a_0(980)^0\to \eta'\pi^0)$       &       $0.04\pm0.01$    & &      $0.05\pm0.02$      \\
$\mathcal{B}(a_0(980)^0\to K^+K^-)$       &       $0.19\pm0.02$        & &     $0.15\pm0.03$      \\
$\mathcal{B}(a_0(980)^0\to K^0\bar{K}^0)$       &       $0.17\pm0.01$  & &      $0.13\pm0.03$    \\\hline
                                          &       $0.45\pm0.09_{\theta_S=[25^\circ,40^\circ]}$         & $0.43\pm0.07_{\theta_S=[25^\circ,35^\circ]}$                                             &  $0.42\pm0.16 ^\dag$    \\
$\mathcal{B}(f_0(980)\to \pi^+\pi^-)$      &        $0.36\pm0.17_{\theta_S=[140^\circ,165^\circ]}$     & $0.41\pm0.09_{\theta_S=[144^\circ,158^\circ]}$                                              &  $0.59\pm0.13^\sharp$    \\
                                          &       $0.22\pm0.22_{\theta_S=[-30^\circ,30^\circ]}$        &$0.38\pm0.06_{[22^\circ\leq|\theta_S|\leq30^\circ]}$ &                             \\ \hline
                                           &       $0.22\pm0.04_{\theta_S=[25^\circ,40^\circ]}$        &$0.21\pm0.03_{\theta_S=[25^\circ,35^\circ]}$                                              &  $0.34\pm0.11 ^\dag$    \\
$\mathcal{B}(f_0(980)\to \pi^0\pi^0)$      &       $0.18\pm0.09_{\theta_S=[140^\circ,165^\circ]}$      &$0.21\pm0.04_{\theta_S=[144^\circ,158^\circ]}$                                              &  $0.20\pm0.10^\sharp$    \\
                                           &       $0.11\pm0.11_{\theta_S=[-30^\circ,30^\circ]}$       &$0.19\pm0.03_{[22^\circ\leq|\theta_S|\leq30^\circ]}$ &                            \\\hline
                                          &       $0.17\pm0.07_{\theta_S=[25^\circ,40^\circ]}$         &$0.19\pm0.05_{\theta_S=[25^\circ,35^\circ]}$                                               &                     \\
$\mathcal{B}(f_0(980)\to K^+K^-)$         &       $0.24\pm0.14_{\theta_S=[140^\circ,165^\circ]}$       &$0.20\pm0.07_{\theta_S=[144^\circ,158^\circ]}$                                               &  $0.12\pm0.04$      \\
                                          &       $0.35\pm0.17_{\theta_S=[-30^\circ,30^\circ]}$        &$0.22\pm0.04_{[22^\circ\leq|\theta_S|\leq30^\circ]}$ &                        \\\hline
                                           &       $0.16\pm0.06_{\theta_S=[25^\circ,40^\circ]}$        &$0.17\pm0.05_{\theta_S=[25^\circ,35^\circ]}$                                              &                     \\
$\mathcal{B}(f_0(980)\to K^0\bar{K}^0)$    &       $0.22\pm0.12_{\theta_S=[140^\circ,165^\circ]}$      &$0.18\pm0.06_{\theta_S=[144^\circ,158^\circ]}$                                               & $0.11\pm0.04$     \\
                                           &       $0.32\pm0.16_{\theta_S=[-30^\circ,30^\circ]}$       &$0.20\pm0.04_{[22^\circ\leq|\theta_S|\leq30^\circ]}$ &                    \\\hline
$\mathcal{B}(f_0(500)\to \pi^+\pi^-)$      &       $0.66\pm0.00$          & &  $0.73\pm0.09^\dag$     \\
                                            &                             & &  $0.57\pm0.12^\sharp$     \\\hline
$\mathcal{B}(f_0(500)\to \pi^0\pi^0)$       &       $0.34\pm0.00$         & &  $0.27\pm0.09^\dag$    \\
                                            &                             & &  $0.43\pm0.12^\sharp$    \\\hline
\end{tabular}\label{Tab:BrS2PP}
\end{table}

\begin{sidewaystable}
\renewcommand\arraystretch{1.2}
\tabcolsep 0.12in
\centering
\caption{ The experimental data and the SU(3) flavor symmetry  predictions
 of the $D\to S(S\to P_1P_2)\ell^+\nu_\ell$ decays with the $c \to s\ell^+\nu_\ell$ transitions within $2\sigma$ errors.  $^\dag$denotes the results  with $\frac{g'_4}{g_4}>0$, and $^\sharp$ denotes ones with $\frac{g'_4}{g_4}<0$.} \vspace{0.1cm}
\begin{tabular}{l|c|ccc|c}  \hline
 Branching ratios   & Exp. Data           &  \multicolumn{3}{c|}{Ones in the 2-quark picture with}   & Ones in the 4-quark picture    \\
  & &$\theta_S=[25^\circ,35^\circ]$  & $\theta_S=[144^\circ,158^\circ]$ & $22^\circ\leq|\theta_S|\leq30^\circ$ & \\\hline
$\mathcal{B}(D^0\to K^-_0e^+\nu_e,~K^-_0\to \pi^-\overline{K}^0)(\times10^{-4})$& $\cdots$                                           & $19.99\pm7.34$      & $19.86\pm7.26$   & $19.74\pm6.97$    &   $8.37\pm3.01$    \\
$\mathcal{B}(D^0\to K^-_0e^+\nu_e,~K^-_0\to \pi^0K^-)(\times10^{-4})$& $\cdots$                                                      & $10.18\pm3.77$      & $10.12\pm3.73$   & $10.05\pm3.57$    &   $4.19\pm1.50$    \\
$\mathcal{B}(D^+\to \overline{K}^0_0e^+\nu_e,~\overline{K}^0_0\to \pi^+K^-)(\times10^{-3})$& $\cdots$                                & $5.17\pm1.92$       & $5.19\pm1.85$    &  $5.12\pm1.86$   &   $2.24\pm0.83$    \\
$\mathcal{B}(D^+\to \overline{K}^0_0e^+\nu_e,~\overline{K}^0_0\to \pi^0\overline{K}^0)(\times10^{-3})$&$\cdots$                      & $2.57\pm0.96$       & $2.59\pm0.92$       &  $2.55\pm0.92$   &   $1.12\pm0.42$    \\
$\mathcal{B}(D^+_s\to f_0(980)e^+\nu_e,~f_0(980)\to \pi^+\pi^-)(\times10^{-3})$& $1.30\pm0.63$  \cite{Hietala:2015jqa}         & $1.19\pm0.18$       &$1.17\pm0.17$     &  $1.18\pm0.17$   &   $1.22\pm0.55^\dag,~~1.44\pm0.49^\sharp$    \\
$\mathcal{B}(D^+_s\to f_0(980)e^+\nu_e,~f_0(980)\to \pi^0\pi^0)(\times10^{-4})$& $7.9\pm2.9$ \cite{BESIII:2021pdt}             & $5.95\pm0.92$       &$5.89\pm0.85$     &  $5.90\pm0.86$   &   $7.91\pm2.85^\dag,~~7.13\pm2.10^\sharp$    \\
$\mathcal{B}(D^+_s\to f_0(980)e^+\nu_e,~f_0(980)\to K^+K^-)(\times10^{-4})$& $\cdots$                                                & $5.11\pm2.34$       & $5.53\pm2.78$    &  $6.28\pm2.07$   &   $3.33\pm1.53^\dag,~~3.07\pm1.34^\sharp$    \\
$\mathcal{B}(D^+_s\to f_0(980)e^+\nu_e,~f_0(980)\to K^0\overline{K}^0)(\times10^{-4})$& $\cdots$                                     & $4.62\pm2.12$       &$5.01\pm2.52$     &  $5.68\pm1.87$   &    $3.01\pm1.39^\dag,~~2.78\pm1.22^\sharp$    \\
$\mathcal{B}(D^+_s\to f_0(500)e^+\nu_e,~f_0(500)\to \pi^+\pi^-)(\times10^{-4})$& $\cdots$                                            & $9.91\pm2.83$       & $9.67\pm3.07$    &  $9.44\pm3.30$   &    $2.49\pm2.49^\dag,~~0.90\pm0.90^\sharp$    \\
$\mathcal{B}(D^+_s\to f_0(500)e^+\nu_e,~f_0(500)\to \pi^0\pi^0)(\times10^{-5})$& $<64$ \cite{BESIII:2021pdt}                   & $49.77\pm14.23$     &$48.57\pm15.43$   &  $47.44\pm16.56$   &    $6.66\pm6.66^\dag,~~0.78\pm0.78^\sharp$\\\hline
$\mathcal{B}(D^0\to K^-_0\mu^+\nu_\mu,~K^-_0\to \pi^-K^0)(\times10^{-4})$& $\cdots$                                                 & $17.27\pm6.48$      & $17.16\pm6.41$   &  $17.04\pm6.14$   &    $7.19\pm2.63$    \\
$\mathcal{B}(D^0\to K^-_0\mu^+\nu_\mu,~K^-_0\to \pi^0K^-)(\times10^{-4})$& $\cdots$                                                 & $8.63\pm3.24$       &  $8.58\pm3.20$   &  $8.52\pm3.07$   &    $3.59\pm1.32$    \\
$\mathcal{B}(D^+\to \overline{K}^0_0\mu^+\nu_\mu,~\overline{K}^0_0\to \pi^+K^-)(\times10^{-3})$& $\cdots$                           & $4.43\pm1.68$       &$4.46\pm1.62$     &  $4.40\pm1.62$   &    $1.92\pm0.73$    \\
$\mathcal{B}(D^+\to \overline{K}^0_0\mu^+\nu_\mu,~\overline{K}^0_0\to \pi^0K^0)(\times10^{-3})$& $\cdots$                           & $2.22\pm0.84$       & $2.23\pm0.81$    &  $2.20\pm0.81$   &    $0.96\pm0.36$    \\
$\mathcal{B}(D^+_s\to f_0(980)\mu^+\nu_\mu,~f_0(980)\to \pi^+\pi^-)(\times10^{-3})$& $\cdots$                                       & $1.01\pm0.16$       &  $1.00\pm0.15$   &   $1.00\pm0.16$  &    $1.02\pm0.46^\dag,~~1.23\pm0.42^\sharp$    \\
$\mathcal{B}(D^+_s\to f_0(980)\mu^+\nu_\mu,~f_0(980)\to \pi^0\pi^0)(\times10^{-4})$& $\cdots$                                       & $5.05\pm0.83$       &$4.99\pm0.77$     &  $5.00\pm0.78$   &    $6.72\pm2.48^\dag,~~6.04\pm1.82^\sharp$    \\
$\mathcal{B}(D^+_s\to f_0(980)\mu^+\nu_\mu,~f_0(980)\to K^+K^-)(\times10^{-4})$& $\cdots$                                           & $4.31\pm1.94$       &  $4.70\pm2.34$   &   $5.34\pm1.75$  &    $2.79\pm1.28^\dag,~~2.59\pm1.14^\sharp$    \\
$\mathcal{B}(D^+_s\to f_0(980)\mu^+\nu_\mu,~f_0(980)\to K^0\overline{K}^0)(\times10^{-4})$& $\cdots$                                & $3.90\pm1.76$       &$4.25\pm2.12$     &   $4.83\pm1.58$  &    $2.52\pm1.16^\dag,~~2.34\pm1.03^\sharp$    \\
$\mathcal{B}(D^+_s\to f_0(500)\mu^+\nu_\mu,~f_0(500)\to \pi^+\pi^-)(\times10^{-4})$& $\cdots$                                       & $8.88\pm2.62$       &$8.70\pm2.86$     &   $8.49\pm3.05$  &    $2.30\pm2.30^\dag,~~0.83\pm0.83^\sharp$    \\
$\mathcal{B}(D^+_s\to f_0(500)\mu^+\nu_\mu,~f_0(500)\to \pi^0\pi^0)(\times10^{-5})$& $\cdots$                                       & $44.67\pm13.23$     & $43.85\pm14.53$  &  $42.77\pm15.49$   &    $6.16\pm6.16^\dag,~~7.23\pm7.23^\sharp$    \\
\hline
\end{tabular}\label{Tab:D2SlvS2PPcs}
\end{sidewaystable}
\begin{sidewaystable}
\renewcommand\arraystretch{0.95}
\tabcolsep 0.15in
\centering
\caption{ The experimental data and the SU(3) flavor symmetry  predictions
 of the $D\to S(S\to P_1P_2)\ell^+\nu_\ell$ decays with the $c \to d\ell^+\nu_\ell$ transitions within $2\sigma$ errors. $^\dag$ denotes the results  with $\frac{g'_4}{g_4}>0$, $^\sharp$ denotes ones with $\frac{g'_4}{g_4}<0$, and $^a$ denotes the experimental lower limits  not used to constrain the predictions.}
{\footnotesize
\begin{tabular}{l|c|ccc|c}  \hline
 Branching ratios   & Exp. Data           &  \multicolumn{3}{c|}{Ones in the 2-quark picture with}   & Ones in the 4-quark picture    \\
  & &$\theta_S=[25^\circ,35^\circ]$  & $\theta_S=[144^\circ,158^\circ]$ & $22^\circ\leq|\theta_S|\leq30^\circ$ & \\\hline
$\mathcal{B}(D^0\to a_0(980)^-e^+\nu_e,~a_0(980)^-\to \eta\pi^-)(\times10^{-5})$&  $13.3^{+6.8}_{-6.0^a}$    &$5.99\pm2.69$    & $5.86\pm2.48$     &$6.05\pm2.57$     & $3.81\pm0.98$\\
$\mathcal{B}(D^0\to a_0(980)^-e^+\nu_e,~a_0(980)^-\to \eta'\pi^-)(\times10^{-6})$& $\cdots$                        &$2.88\pm1.71$    & $2.97\pm1.77$     &$2.97\pm1.73$     & $1.88\pm0.98$\\
$\mathcal{B}(D^0\to a_0(980)^-e^+\nu_e,~a_0(980)^-\to K^0K^-)(\times10^{-6})$& $\cdots$                            &$29.99\pm13.81$  & $30.73\pm13.81$   & $30.57\pm13.70$  & $4.22\pm1.93$\\
$\mathcal{B}(D^+\to a_0(980)^0e^+\nu_e,~a_0(980)^0\to \eta\pi^0)(\times10^{-5})$& $17^{+16}_{-14}$           &$7.35\pm3.28$    & $7.25\pm3.13$     &$7.32\pm3.17$     & $4.00\pm1.00$\\
$\mathcal{B}(D^+\to a_0(980)^0e^+\nu_e,~a_0(980)^0\to \eta'\pi^0)(\times10^{-6})$& $\cdots$                        &$5.53\pm3.26$    & $5.69\pm3.32$     &$5.65\pm3.20$     & $3.08\pm1.56$\\
$\mathcal{B}(D^+\to a_0(980)^0e^+\nu_e,~a_0(980)^0\to K^+K^-)(\times10^{-5})$& $\cdots$                            &$2.28\pm1.06$    & $2.30\pm1.00$     &$2.29\pm0.99$     & $0.88\pm0.36$\\
$\mathcal{B}(D^+\to a_0(980)^0e^+\nu_e,~a_0(980)^0\to K^0\overline{K}^0)(\times10^{-5})$& $\cdots$                 &$1.99\pm0.92$    & $2.01\pm0.88$     &$2.00\pm0.86$     & $0.77\pm0.31$\\
$\mathcal{B}(D^+\to f_0(980)e^+\nu_e,~f_0(980)\to \pi^+\pi^-)(\times10^{-5})$& $<2.8~\mbox{\cite{BESIII:2018qmf}}$  &$1.15\pm0.50$    & $1.10\pm0.58$     &$0.96\pm0.43$     & $1.65\pm1.15^\dag,~~2.14\pm0.65^\sharp$  \\
$\mathcal{B}(D^+\to f_0(980)e^+\nu_e,~f_0(980)\to \pi^0\pi^0)(\times10^{-6})$& $\cdots$                            &$5.75\pm2.53$    & $5.51\pm2.92$     &$4.80\pm2.18$     & $10.53\pm3.67^\dag,~~10.10\pm5.37^\sharp$   \\
$\mathcal{B}(D^+\to f_0(980)e^+\nu_e,~f_0(980)\to K^+K^-)(\times10^{-6})$& $\cdots$                                &$5.07\pm0.88$    &$5.06\pm0.85$      &$5.01\pm0.80$     & $4.35\pm2.78^\dag,~~4.60\pm2.76^\sharp$  \\
$\mathcal{B}(D^+\to f_0(980)e^+\nu_e,~f_0(980)\to K^0\overline{K}^0)(\times10^{-6})$& $\cdots$                     &$5.07\pm0.88$    & $5.06\pm0.85$     &$5.01\pm0.80$     & $4.35\pm2.78^\dag,~~4.60\pm2.76^\sharp$ \\
$\mathcal{B}(D^+\to f_0(500)e^+\nu_e,~f_0(500)\to \pi^+\pi^-)(\times10^{-4})$& $6.3\pm1.0^a$                 &$1.44\pm0.64$    & $1.72\pm0.92$     &$1.79\pm0.85$     & $3.64\pm2.57^\dag,~~2.95\pm1.87^\sharp$ \\
$\mathcal{B}(D^+\to f_0(500)e^+\nu_e,~f_0(500)\to \pi^0\pi^0)(\times10^{-4})$& $\cdots$                            &$0.72\pm0.32$    & $0.87\pm0.46$     &$0.91\pm0.43$     & $1.45\pm1.02^\dag,~~2.08\pm1.57^\sharp$   \\
$\mathcal{B}(D^+_s\to K^0_0e^+\nu_e,~K^0_0\to \pi^-K^+)(\times10^{-5})$& $\cdots$                                  &$22.34\pm8.09$   & $22.13\pm7.97$    & $22.34\pm7.64$   & $9.54\pm3.38$\\
$\mathcal{B}(D^+_s\to K^0_0e^+\nu_e,~K^0_0\to \pi^0K^0)(\times10^{-5})$& $\cdots$                                  &$11.17\pm4.04$   & $11.07\pm3.99$    & $11.17\pm3.82$   & $4.77\pm1.69$\\
\hline
$\mathcal{B}(D^0\to a_0(980)^-\mu^+\nu_\mu,~a_0(980)^-\to \eta\pi^-)(\times10^{-5})$& $\cdots$                     &$4.95\pm2.27$    & $4.84\pm2.10$     &$5.00\pm2.18$     & $3.14\pm0.84$\\
$\mathcal{B}(D^0\to a_0(980)^-\mu^+\nu_\mu,~a_0(980)^-\to \eta'\pi^-)(\times10^{-6})$& $\cdots$                    &$2.39\pm1.44$    & $2.46\pm1.48$     &$2.45\pm1.45$     & $1.56\pm0.82$\\
$\mathcal{B}(D^0\to a_0(980)^-\mu^+\nu_\mu,~a_0(980)^-\to K^0K^-)(\times10^{-6})$& $\cdots$                        &$24.78\pm11.68$  & $25.37\pm11.62$   & $25.20\pm11.53$  & $3.51\pm1.62$\\
$\mathcal{B}(D^+\to a_0(980)^0\mu^+\nu_\mu,~a_0(980)^0\to \eta\pi^0)(\times10^{-5})$& $\cdots$                     &$6.09\pm2.78$    & $6.00\pm2.65$     &$6.06\pm2.69$     & $3.30\pm0.86$\\
$\mathcal{B}(D^+\to a_0(980)^0\mu^+\nu_\mu,~a_0(980)^0\to \eta'\pi^0)(\times10^{-6})$& $\cdots$                    &$4.58\pm2.74$    & $4.72\pm2.79$     &$4.67\pm2.69$     & $2.55\pm1.31$\\
$\mathcal{B}(D^+\to a_0(980)^0\mu^+\nu_\mu,~a_0(980)^0\to K^+K^-)(\times10^{-5})$& $\cdots$                        &$1.89\pm0.89$    & $1.91\pm0.85$     &$1.89\pm0.83$     & $0.73\pm0.30$\\
$\mathcal{B}(D^+\to a_0(980)^0\mu^+\nu_\mu,~a_0(980)^0\to K^0\overline{K}^0)(\times10^{-5})$& $\cdots$             &$1.65\pm0.78$    & $1.66\pm0.74$     &$1.65\pm0.73$     & $0.64\pm0.27$\\
$\mathcal{B}(D^+\to f_0(980)\mu^+\nu_\mu,~f_0(980)\to \pi^+\pi^-)(\times10^{-5})$& $\cdots$                        &$0.94\pm0.43$    & $0.91\pm0.48$     &$0.79\pm0.36$     & $1.37\pm0.96^\dag,~~1.76\pm0.55^\sharp$    \\
$\mathcal{B}(D^+\to f_0(980)\mu^+\nu_\mu,~f_0(980)\to \pi^0\pi^0)(\times10^{-6})$& $\cdots$                        &$4.74\pm2.14$    & $4.58\pm2.43$     &$3.97\pm1.82$     & $8.67\pm3.13^\dag,~~8.32\pm4.47^\sharp$      \\
$\mathcal{B}(D^+\to f_0(980)\mu^+\nu_\mu,~f_0(980)\to K^+K^-)(\times10^{-6})$& $\cdots$                            &$4.21\pm0.73$    & $4.19\pm0.71$     &$4.15\pm0.67$     &  $3.55\pm2.29^\dag,~~3.76\pm2.26^\sharp$         \\
$\mathcal{B}(D^+\to f_0(980)\mu^+\nu_\mu,~f_0(980)\to K^0\overline{K}^0)(\times10^{-6})$& $\cdots$                 &$4.21\pm0.73$    & $4.19\pm0.71$     &$4.15\pm0.67$     &  $3.55\pm2.29^\dag,~~3.76\pm2.26^\sharp$   \\
$\mathcal{B}(D^+\to f_0(500)\mu^+\nu_\mu,~f_0(980)\to \pi^+\pi^-)(\times10^{-4})$& $\cdots$                        &$1.28\pm0.59$    & $1.54\pm0.84$     &$1.61\pm0.79$     & $3.30\pm2.39^\dag,~~2.68\pm1.74^\sharp$  \\
$\mathcal{B}(D^+\to f_0(500)\mu^+\nu_\mu,~f_0(980)\to \pi^0\pi^0)(\times10^{-4})$& $\cdots$                        &$0.64\pm0.30$    & $0.78\pm0.43$     &$0.81\pm0.40$     & $1.32\pm0.95^\dag,~~1.89\pm1.46^\sharp$   \\
$\mathcal{B}(D^+_s\to K^0_0\mu^+\nu_\mu,~K^0_0\to \pi^-K^+)(\times10^{-5})$& $\cdots$                              &$19.61\pm7.20$   & $19.43\pm7.10$    &$19.60\pm6.80$    & $8.38\pm3.01$\\
$\mathcal{B}(D^+_s\to K^0_0\mu^+\nu_\mu,~K^0_0\to \pi^0K^0)(\times10^{-5})$& $\cdots$                              &$9.80\pm3.60$    & $9.71\pm3.55$     &$9.80\pm3.40$     & $4.19\pm1.50$\\
\hline
\end{tabular} }\label{Tab:D2SlvS2PPcd}
\end{sidewaystable}


\subsection{$D\to V(V\to P_1P_2) \ell^+\nu_\ell$ decays} \label{Sec:V}

We will analyze the  $D\to P_1P_2\ell^+\nu_\ell$  decays with the vector resonances in this subsection.  Since the light vector mesons are understood well,  the calculations of $\mathcal{B}(D\to V\ell^+\nu_\ell,V\to P_1P_2)$  are much easier than the ones of $\mathcal{B}(D\to S\ell^+\nu_\ell,S\to P_1P_2)$.   From  Eq. (\ref{Eq:BrD2RlvR2PP}), their branching ratios of $D\to V(V\to P_1P_2) \ell^+\nu_\ell$ can be obtained by using $\mathcal{B}(D\to V\ell^+\nu_\ell)$ and $\mathcal{B}(V\to P_1P_2)$.
The $D\to V\ell^+\nu_\ell$ decays have been studied  by the SU(3) flavor symmetry in Ref. \cite{Wang:D2MlvSU3}.  Many   $\mathcal{B}(D\to V\ell^+\nu_\ell)$ have been accurately measured   and have been listed in the second column of Tab. V in Ref.  \cite{Wang:D2MlvSU3}. The expressions  of $\mathcal{B}(D\to V\ell^+\nu_\ell)$ within the $C_3$  case in Ref.  \cite{Wang:D2MlvSU3}  will be taken  for our analysis.

Following Ref. \cite{Cheng:2020ipp}, $\mathcal{B}(V\to P_1P_2)$  can be written as
\begin{eqnarray}
\mathcal{B}(V\to P_1P_2)=\frac{\tau_Vp'^3_c}{6\pi m^2_V}g^2_{V\to P_1P_2},
\end{eqnarray}
where $p'_c\equiv\frac{\sqrt{\lambda(m_V^2,m_{P_1}^2,m_{P_2}^2)}}{2m_V}$  and  $g_{V\to P_1P_2}$ are the strong coupling constants.  Similar to $g_{S\to P_1P_2}^{2q}$  in Eq. (\ref{Eq:gS2PP2q}),   $g_{V\to P_1P_2}$  can be  parametrized  by  the SU(3) flavor symmetry
\begin{eqnarray}
g_{V\to P_1P_2}=g_VV^i_jP^k_iP^j_k, \label{Eq:gV2PP0}
\end{eqnarray}
where $g_V$ is the corresponding nonperturbative parameter.

At present, many involved $\mathcal{B}(V\to P_1P_2)$ have been well measured \cite{PDG2022}
\begin{eqnarray}
&\mathcal{B}(K^{*+}\to\pi K)=(99.902\pm0.018)\%,~~~~~~~~~~~~~~&\mathcal{B}(K^{*0}\to\pi K)=(99.754\pm0.042)\%,\nonumber\\
&\mathcal{B}(\rho^{+}\to\pi^0\pi^+)=100\%,~~~~~~~~~~~~~~~~~~~~~~~~~~~~~~~~&\mathcal{B}(\rho^{0}\to\pi^+\pi^-)=100\%,\nonumber\\
&\mathcal{B}(\phi\to K^+K^-)=(49.1\pm1.0)\%,~~~~~~~~~~~~~~~~~~~~~~&\mathcal{B}(\omega\to\pi^+\pi^-)=(1.53^{+0.22}_{-0.26})\%. \label{Eq:BrV2PPE}
\end{eqnarray}
From Eq. (\ref{Eq:gV2PP0}), the  relations of the strong coupling constants can be obtained
\begin{eqnarray}
&&\sqrt{2}g_{K^{*-}\to \pi^0K^-}=g_{K^{*-}\to \pi^-K^0},~~~~~~~~~~~~~~~~~~~~~~~~~~~~~~~~\sqrt{2}g_{K^{*0}\to \pi^0K^0}=g_{K^{*0}\to \pi^-K^+},\nonumber\\
&&g_{\rho^-\to \pi^0\pi^-}=\sqrt{3}g_{\rho^-\to \eta_8\pi^-}=\sqrt{3/2}g_{\rho^-\to \eta_1\pi^-}, ~~~~~~~~~~~~~~g_{\phi\to K^+K^-}=g_{\phi\to K^0\overline{K}^0},  \label{Eq:gV2PP}
\end{eqnarray}
In terms of Eq. (\ref{Eq:BrV2PPE}) and Eq. (\ref{Eq:gV2PP}), the strong coupling constants are
\begin{eqnarray}
&|g_{K^{*-}\to \pi^-K^0}|=4.62\pm0.08,~~~~~~~~~~~~~~&|g_{K^{*0}\to \pi^-K^+}|=4.40\pm0.10,\nonumber\\
&|g_{\rho^-\to \pi^0\pi^-}|=6.00\pm0.03,~~~~~~~~~~~~~~&|g_{\rho^{0}\to\pi^+\pi^-}|=5.95\pm0.04,\nonumber\\
&|g_{\phi\to K^+K^-}|=4.47\pm0.08,~~~~~~~~~~~~~~&|g_{\omega\to\pi^+\pi^-}|=0.18\pm0.02.
\end{eqnarray}
Then the following $\mathcal{B}(V\to P_1P_2)$ can be written as
\begin{eqnarray}
&\mathcal{B}(K^{*0}\to\pi^0 K^0)=(33.02\pm0.02)\%,~~~~~~~~~~~~~~&\mathcal{B}(K^{*0}\to\pi^- K^+)=(66.74\pm0.04)\%,\nonumber\\
&\mathcal{B}(K^{*+}\to\pi^0 K^+)=(33.62\pm0.01)\%,~~~~~~~~~~~~~~&\mathcal{B}(K^{*+}\to\pi^+ K^0)=(66.28\pm0.01)\%,\nonumber\\
&\mathcal{B}(\rho^+\to \eta \pi^+)=(4.38\pm0.66)\%,~~~~~~~~~~~~~~&\mathcal{B}(\phi\to K^0K^0)=(32.42\pm1.04)\%.
\end{eqnarray}

\begin{table}[b]
\renewcommand\arraystretch{1.25}
\tabcolsep 0.15in
\centering
\caption{ The experimental data and the SU(3) flavor symmetry  predictions
 of $D\to V(V\to P_1P_2)\ell^+\nu_\ell$ decays  within $2\sigma$ errors.}
\vspace{0.1cm}
\begin{tabular}{l c cc}  \hline
 Branching ratios   & Exp. Data       &  Our predictions & Previous ones   \\\hline
{\color{blue}$c \to se^+\nu_e$:}&&\\
$\mathcal{B}(D^0\to K^{*-}e^+\nu_e,~K^{*-}\to \pi^-\overline{K}^0)(\times10^{-2})$         &   $ \dots $                      &   $1.42\pm0.07  $&   $ \dots $            \\
$\mathcal{B}(D^0\to K^{*-}e^+\nu_e,~K^{*-}\to \pi^0K^-)(\times10^{-3})$                    &   $ \dots $                      &   $7.18\pm0.37  $  &  $7.17$ \cite{Kim:2017dfr}       \\
$\mathcal{B}(D^+\to \overline{K}^{*0}e^+\nu_e,~\overline{K}^{*0}\to \pi^+K^-)(\times10^{-2})$ &   $ 3.77\pm0.34 $        &   $ 3.64\pm0.11 $  &  $3.51$ \cite{Kim:2017dfr}         \\
$\mathcal{B}(D^+\to \overline{K}^{*0}e^+\nu_e,~\overline{K}^{*0}\to \pi^0\overline{K}^0)(\times10^{-2})$    &   $ \dots $     &   $1.80\pm0.06  $ &   $ \dots $           \\
$\mathcal{B}(D^+_s\to \phi e^+\nu_e,~\phi\to K^+K^-)(\times10^{-2})$                       &   $ \dots $                      &   $ 1.20\pm0.10 $ &   $ \dots $           \\
$\mathcal{B}(D^+_s\to \phi e^+\nu_e,~\phi\to K^0\overline{K}^0)(\times10^{-3})$            &   $ \dots $                      &   $  7.94\pm0.65$  &   $ \dots $          \\
\hline
{\color{blue}$c \to s\mu^+\nu_\mu$:}&&\\
$\mathcal{B}(D^0\to K^{*-}\mu^+\nu_\mu,~K^{*-}\to \pi^-\overline{K}^0)(\times10^{-2})$     &   $ \dots $                       &   $ 1.33\pm0.07 $ &   $ \dots $           \\
$\mathcal{B}(D^0\to K^{*-}\mu^+\nu_\mu,~K^{*-}\to \pi^0K^-)(\times10^{-3})$                &   $ \dots $                       &   $ 6.76\pm0.35 $  &  $7.17$ \cite{Kim:2017dfr}         \\
$\mathcal{B}(D^+\to \overline{K}^{*0}\mu^+\nu_\mu,~\overline{K}^{*0}\to \pi^+K^-)(\times10^{-2})$   &   $ 3.52\pm0.20 $   &   $ 3.43\pm0.11 $  &  $3.51$ \cite{Kim:2017dfr}         \\
$\mathcal{B}(D^+\to \overline{K}^{*0}\mu^+\nu_\mu,~\overline{K}^{*0}\to \pi^0\overline{K}^0)(\times10^{-2})$  &   $\dots  $    &   $ 1.70\pm0.05 $  &   $ \dots $          \\
$\mathcal{B}(D^+_s\to \phi \mu^+\nu_\mu,~\phi\to K^+K^-)(\times10^{-2})$                   &   $\dots  $                       &   $ 1.13\pm0.09 $  &   $ \dots $          \\
$\mathcal{B}(D^+_s\to \phi \mu^+\nu_\mu,~\phi\to K^0\overline{K}^0)(\times10^{-3})$        &   $\dots  $                       &   $ 7.46\pm0.62 $  &   $ \dots $          \\
\hline
{\color{blue}$c \to de^+\nu_e$:}&&\\
$\mathcal{B}(D^0\to \rho^-e^+\nu_e,~\rho^-\to \pi^0\pi^-)(\times10^{-3})$              &   $\dots  $                      &   $ 1.85\pm0.11 $    &  $1.63$ \cite{Kim:2017dfr}       \\
$\mathcal{B}(D^0\to \rho^-e^+\nu_e,~\rho^-\to \eta\pi^-)(\times10^{-5})$               &   $\dots  $                      &   $ 8.23\pm1.59 $ &   $ \dots $           \\
$\mathcal{B}(D^+\to \rho^0e^+\nu_e,~\rho^0\to \pi^+\pi^-)(\times10^{-3})$              &   $\dots  $                      &   $ 2.40\pm0.12 $    & $1.57\pm0.07~\mbox{\cite{Shi:2017pgh}},~~2.10~\mbox{\cite{Kim:2017dfr}}$      \\
$\mathcal{B}(D^+\to \omega e^+\nu_e,~\omega\to \pi^+\pi^-)(\times10^{-5})$             &   $ \dots $                      &   $ 3.55\pm0.82 $   &   $ \dots $         \\
$\mathcal{B}(D^+_s\to K^{*0} e^+\nu_e,~K^{*0}\to \pi^-K^+)(\times10^{-3})$             &   $ \dots $                      &   $ 1.49\pm0.10 $  &   $ \dots $          \\
$\mathcal{B}(D^+_s\to K^{*0} e^+\nu_e,~K^{*0}\to \pi^0K^0)(\times10^{-4})$             &   $ \dots $                      &   $ 7.39\pm0.51 $   &   $ \dots $         \\\hline
{\color{blue}$c \to d\mu^+\nu_\mu$:}&&\\
$\mathcal{B}(D^0\to \rho^-\mu^+\nu_\mu,~\rho^-\to \pi^0\pi^-)(\times10^{-3})$              &   $\dots  $                      &   $ 1.76\pm0.10 $  &   $ \dots $          \\
$\mathcal{B}(D^0\to \rho^-\mu^+\nu_\mu,~\rho^-\to \eta\pi^-)(\times10^{-5})$               &   $\dots  $                      &   $ 7.83\pm1.51 $  &   $ \dots $          \\
$\mathcal{B}(D^+\to \rho^0\mu^+\nu_\mu,~\rho^0\to \pi^+\pi^-)(\times10^{-3})$              &   $\dots  $                      &   $ 2.29\pm0.11 $   & $ 1.57\pm0.07 $ \cite{Shi:2017pgh}          \\
$\mathcal{B}(D^+\to \omega \mu^+\nu_\mu,~\omega\to \pi^+\pi^-)(\times10^{-5})$             &   $\dots  $                      &   $ 3.38\pm0.78 $  &   $ \dots $          \\
$\mathcal{B}(D^+_s\to K^{*0} \mu^+\nu_\mu,~K^{*0}\to \pi^-K^+)(\times10^{-3})$             &   $\dots  $                      &   $ 1.42\pm0.10 $  &   $ \dots $          \\
$\mathcal{B}(D^+_s\to K^{*0} \mu^+\nu_\mu,~K^{*0}\to \pi^0K^0)(\times10^{-4})$             &   $\dots  $                      &   $ 7.03\pm0.48 $  &   $ \dots $          \\\hline
\end{tabular}\label{Tab:BD2PPlvV}
\end{table}
For $D\to V(V\to P_1P_2)\ell^+\nu_\ell$ decays,  the branching ratios of $D^+\to \overline{K}^{*0}(\overline{K}^{*0}\to \pi^+K^-)e^+\nu_e$ and $D^+\to \overline{K}^{*0}(\overline{K}^{*0}\to \pi^+K^-)\mu^+\nu_\mu$ have been measured, and the experimental  data with $2\sigma$ errors are listed in the second column of Tab. \ref{Tab:BD2PPlvV}.  Using the experimental  data of $\mathcal{B}(D^+\to \overline{K}^{*0}\ell^+\nu_\ell,\overline{K}^{*0}\to \pi^+K^-)$, $\mathcal{B}(V\to P_1P_2)$  and  $\mathcal{B}(D\to V\ell^+\nu_\ell)$, we obtain the predictions of $\mathcal{B}(D\to V\ell^+\nu_\ell,V\to P_1P_2)$  by the SU(3) flavor symmetry, which are  given in the third  column of Tab. \ref{Tab:BD2PPlvV}. We can see that   $\mathcal{B}(D\to V\ell^+\nu_\ell,V\to P_1P_2)$ with the $c \to s\ell^+\nu_\ell$ transitions are predicted on the order of $\mathcal{O}(10^{-2}-10^{-3})$, and  $\mathcal{B}(D\to V\ell^+\nu_\ell,V\to P_1P_2)$ with the $c \to d\ell^+\nu_\ell$ transitions are predicted on the order of $\mathcal{O}(10^{-3}-10^{-5})$.  The predictions of $\mathcal{B}(D\to V\ell^+\nu_\ell,V\to P_1P_2)$ are about one order larger than those of the corresponding $\mathcal{B}(D\to S\ell^+\nu_\ell,S\to P_1P_2)$.

Previous predictions are also listed in the last column of   Tab. \ref{Tab:BD2PPlvV}.  Our predictions of $\mathcal{B}(D^0\to K^{*-}\ell^+\nu_\ell,~K^{*-}\to \pi^0K^-)$ and $\mathcal{B}(D^+\to \overline{K}^{*0}\ell^+\nu_\ell,~\overline{K}^{*0}\to \pi^+K^-)$ are in good  agreement with those in  Ref. \cite{Kim:2017dfr}.
And our predictions of $\mathcal{B}(D^+\to \rho^0\ell^+\nu_\ell,~\rho^0\to \pi^+\pi^-)$ are   slight larger than those  obtained by the light-front quark model and the light-cone sum rules in Ref. \cite{Shi:2017pgh}.


\subsection{Total branching ratios}
As analyzed  above, some four-body semileptonic decays of $D$ mesons receive the contributions of the nonresonant states, the scalar resonant states, and the  vector resonant states; nevertheless,  some  decay modes only receive one or two kinds of  them.  For clearly showing the  resonant contributions, we also list the scalar and vector resonant amplitudes in the third and last columns of Tab. \ref{Tab:HD2PPlvAmp}, respectively. The resonant amplitudes are obtained by multiplying the hadronic helicity amplitudes $H(D\to R\ell^+\nu_\ell)$  given in Ref. \cite{Wang:D2MlvSU3} and the strong coupling constants  $g_{R\to P_1P_2}$  obtained in this work.  Note that the resonant amplitudes listed in the last two columns of Tab. \ref{Tab:HD2PPlvAmp} are given only to see clearly  the kinds of the resonant contributions, and we do not use them to obtain the numerical total branching ratios $\mathcal{B}(D\to P_1P_2\ell^+\nu_\ell)_T$.

We have some comments for the contributions in  Tab. \ref{Tab:HD2PPlvAmp}.
For $D_{(s)}\to \eta K\ell^+\nu_\ell,~\eta'K\ell^+\nu_\ell,~\eta\eta\ell^+\nu_\ell,~\eta\eta'\ell^+\nu_\ell$ decays, since   both final state mesons  are quite heavy,  they only receive the nonresonant contributions.
The decays  $D_s^+\to \pi^0\pi^0\ell^+\nu_\ell,$ $D_s^+\to\pi^+\pi^-\ell^+\nu_\ell,$  $D^0\to K^-K^0\ell^+\nu_\ell,$ $D^+\to \overline{K}^0K^0\ell^+\nu_\ell,$ $D^+\to K^+K^-\ell^+\nu_\ell,$  $D^+\to \pi^0\pi^0\ell^+\nu_\ell$,  and  $D^+\to \eta^{(')}\pi^0\ell^+\nu_\ell$ receive both  the nonresonant contributions and the scalar resonant contributions;  moreover,
 the nonresonant  contributions in the $D_s^+\to \pi^0\pi^0\ell^+\nu_\ell,$ $D_s^+\to\pi^+\pi^-\ell^+\nu_\ell$ and $D^+\to K^+K^-\ell^+\nu_\ell$ decays are suppressed by the OZI rule,   and the main contributions of these decay branching ratios come from the scalar resonant states.
All other decay modes except the $D^0\to \pi^0\pi^-\ell^+\nu_\ell$  decays  receive all three kinds of the contributions,  and their branching ratios  are dominant by the   vector resonant states. Due to the quantum number constraint, the $D^0\to \pi^0\pi^-\ell^+\nu_\ell$ decays only receive  the contributions of  the  vector resonant states.

In the last columns of Tabs. \ref{Tab:BrD2MMlvcs} and \ref{Tab:BrD2MMlvcd}, total branching ratio predictions  of the $D\to P_1P_2\ell^+\nu$ decays  including the possible nonresonant, scalar resonant and  vector resonant contributions   are listed. The present six experimental data with $2\sigma$ errors are also listed in the forth column of Tab. \ref{Tab:BrD2MMlvcs} and in third  column of Tab. \ref{Tab:BrD2MMlvcd} for convenient comparison.  One can see that  for $\mathcal{B}(D^0\to \pi^-\overline{K}^-e^+\nu_e)$, $\mathcal{B}(D^0\to \pi^0K^-e^+\nu_e)$, $\mathcal{B}(D^+\to \pi^+K^-e^+\nu_e)$, $\mathcal{B}(D^+\to \pi^+K^-\mu^+\nu_\mu)$, and $\mathcal{B}(D^+\to \pi^+\pi^-e^+\nu_e)$,  our SU(3) flavor symmetry predictions are consistent with present data within $2\sigma$ error bars. Our prediction of $\mathcal{B}(D^0\to \pi^0\pi^-e^+\nu_e)$ is slightly larger than its experimental datum; nevertheless, the prediction will be very close to the  datum  within $3\sigma$ error bars.

For some Cabibbo  suppressed decays due to $c\to d\ell^+\nu_\ell$ transitions, such as  the $D^0\to K^-K^0\ell^+\nu_\ell,$  $D^0\to \eta'\pi^-\ell^+\nu_\ell,$  $D^+\to \overline{K}^0K^0\ell^+\nu_\ell,$ $D^+\to \pi^0\pi^0\ell^+\nu_\ell$,  $D^+\to \eta\pi^0\ell^+\nu_\ell$ and  $D^+\to \eta'\pi^0\ell^+\nu_\ell$ decays, they  only receive  both  the nonresonant contributions and the scalar resonant contributions, and we can see that both  the nonresonant   and the scalar resonant contributions are  important.  The  nonresonant contributions in the $D^+\to K^+K^-\ell^+\nu_\ell$ decays are suppressed by the OZI rule,   and the scalar resonant contributions in the $D^+\to K^+K^-\ell^+\nu_\ell$ decays are dominant.

Please note that  the interference terms between nonresonant, scalar,  and vector resonant contributions  exist.  As discussed in Refs. \cite{Kang:2013jaa, Faller:2013dwa},  the interference terms between different partial waves
vanish upon angular integration in the branching ratios, but they may effect
 a number of angular observables of these decays, which have not been discussed in this work.  Nevertheless,  there still are the interference effects between nonresonant and resonant contributions  as well as the ones between different scalar resonances in the $D\to P_1P_2\ell^+\nu_\ell$  decays, for example, between  $D^+\to (a_0(980)\to \overline{K}^0K^0)\ell^+\nu_\ell$ and $D^+\to (f_0(980)\to\overline{K}^0K^0)\ell^+\nu_\ell$. So the interference effects  might also be important for the $D^0\to \pi^0K^-\ell^+\nu_\ell,$ $D^+_s\to \pi^+\pi^-\ell^+\nu_\ell,$ $D^+_s\to \pi^0\pi^0\ell^+\nu_\ell,$ $ D^0\to K^-K^0\ell^+\nu_\ell,$  $D^0\to \eta\pi^-\ell^+\nu_\ell,$  $D^0\to \eta'\pi^-\ell^+\nu_\ell,$  $D^+\to \overline{K}^0K^0\ell^+\nu_\ell,$ $D^+\to \pi^+\pi^-\ell^+\nu_\ell$, $D^+\to \pi^0\pi^0\ell^+\nu_\ell$,  $D^+\to \eta\pi^0\ell^+\nu_\ell$ and  $D^+\to \eta'\pi^0\ell^+\nu_\ell$ decays, in which the two or three kinds of contributions are important.  Currently, we  cannot determine the  size of interference effects by the SU(3) flavor symmetry.

\section{Summary}
Semileptonic decays of heavy mesons are quite interesting   not only  because of relatively simple theoretical description but also the clean experimental signals.
Some semileptonic decays $D\to P_1P_2\ell^+\nu_\ell$ have been measured by  BESIII, CLEO, and $BABAR$, etc.
Using the present data of $\mathcal{B}(D\to P_1P_2\ell^+\nu_\ell)$ and the SU(3) flavor symmetry, we have presented a theoretical analysis of the $D\to P_1P_2\ell^+\nu_\ell$ decays with the nonresonant, the light scalar meson resonant, and the vector meson resonant contributions.

\begin{itemize}
\item {\bf Nonresonant $D\to P_1P_2\ell^+\nu_\ell$ decays:} The amplitude relations including  the SU(3)
              flavor  breaking effects  have been obtained.  Almost all amplitudes can be related  after ignoring the OZI suppressed  and the SU(3)
              flavor  breaking contributions. Via the experimental data of the nonresonant branching ratios $\mathcal{B}(D^+\to \pi^+K^-\ell^+\nu_\ell)_N$, we have predicted other nonresonant branching ratios.  We have found that
              the branching ratios of the nonresonant decays $D^0\to \pi^-\overline{K}^0\ell^+\nu_\ell,\pi^0K^-\ell^+\nu_\ell$, $D^+\to \pi^+K^-\ell^+\nu_\ell,\pi^0\overline{K}^0\ell^+\nu_\ell,\pi^+\pi^-\ell^+\nu_\ell,\pi^0\pi^0\ell^+\nu_\ell$, and $D^+_s\to K^+K^-\ell^+\nu_\ell,K^0\overline{K}^0\ell^+\nu_\ell$   are on the order of $\mathcal{O}(10^{-3}-10^{-4})$, which might be measured by the BESIII, LHCb, and Belle II experiments, and some other decays  might be measured at these experiments in the near future.

\item {\bf Decays with the light scalar meson resonances:} Using the SU(3)
           flavor symmetry and  the present  experimental data of $\mathcal{B}(D\to S\ell^+\nu_\ell)$, $\mathcal{B}(D\to S\ell^+\nu_\ell,S\to P_1P_2)$ as well as $\mathcal{B}(S\to P_1P_2)$, the not-measured $\mathcal{B}(D\to S\ell^+\nu_\ell,S\to P_1P_2)$ have been obtained by  the SU(3) flavor symmetry. We have found that $\mathcal{B}(D\to S\ell^+\nu_\ell,S\to P_1P_2)$ with the $c \to s\ell^+\nu_\ell$ transitions are predicted on the order of $\mathcal{O}(10^{-3}-10^{-4})$,  and $\mathcal{B}(D\to S\ell^+\nu_\ell,S\to P_1P_2)$ with the $c \to d\ell^+\nu_\ell$ transitions are predicted on the order of $\mathcal{O}(10^{-4}-10^{-6})$.
           The two-quark picture and the four-quark picture for the scalar mesons  have been analyzed in the $D\to S(S\to P_1P_2)\ell^+\nu_\ell$ decays.  Present experimental data might favor  the  four-quark picture for the scalar mesons.

\item {\bf Decays with the vector meson resonances: }  Using the experimental data of $\mathcal{B}(D^+\to \overline{K}^{*0}e^+\nu_e,~\overline{K}^{*0}\to \pi^+K^-)$,  $\mathcal{B}(D^+\to \overline{K}^{*0}\mu^+\nu_\mu,~\overline{K}^{*0}\to \pi^+K^-)$, many $\mathcal{B}(D\to V\ell^+\nu_\ell)$ and many $\mathcal{B}(V\to P_1P_2)$,  the not-measured  $B(D\to V\ell^+\nu_\ell,V\to P_1P_2)$ have been predicted by the SU(3) flavor symmetry.    We have found that $\mathcal{B}(D\to V\ell^+\nu_\ell,V\to P_1P_2)$ with the $c \to s\ell^+\nu_\ell$ transitions are predicted on the order of $\mathcal{O}(10^{-2}-10^{-3})$, and  $\mathcal{B}(D\to V\ell^+\nu_\ell,V\to P_1P_2)$ with the $c \to d\ell^+\nu_\ell$ transitions are predicted on the order of $\mathcal{O}(10^{-3}-10^{-5})$.

\item {\bf Total branching ratios:}  Total branching ratio predictions including the possible nonresonant,  light scalar meson resonant and  vector meson resonant contributions have been obtained.
                                  The six total branching ratios have been measured, and we did not use them to further constrain the predictions.   Our five predictions  are consistent with present data within $2\sigma$ errors, and the prediction of $\mathcal{B}(D^0\to \pi^0\pi^-e^+\nu_e)$ will be very close to the  datum  within $3\sigma$ error bars.  We have found that the vector meson resonant contributions are dominant in the $D^0\to \pi^-\overline{K}^0\ell^+\nu_\ell,\pi^0K^-\ell^+\nu_\ell,\pi^0\pi^-\ell^+\nu_\ell$, $D^+\to \pi^+K^-\ell^+\nu_\ell,\pi^0\overline{K}^0\ell^+\nu_\ell,\pi^+\pi^-\ell^+\nu_\ell$, and $D^+_s\to K^+K^-\ell^+\nu_\ell,K^0\overline{K}^0\ell^+\nu_\ell, K^+\pi^-\ell^+\nu_\ell, K^0\pi^0\ell^+\nu_\ell$ decays.  All three kinds of  contributions are   important in $D^0\to\eta\pi^-\ell^+\nu_\ell$ decays. Both the nonresonant and the scalar resonant contributions  are  important in  $D^0\to K^-K^0\ell^+\nu_\ell,\eta'\pi^-\ell^+\nu_\ell$ and $D^+\to \overline{K}^0K^0\ell^+\nu_\ell,\pi^0\pi^0\ell^+\nu_\ell,\eta\pi^0\ell^+\nu_\ell,\eta'\pi^0\ell^+\nu_\ell$ decays.

\end{itemize}

Although SU(3) flavor  symmetry is approximate, it can
still provide  very useful information about these
decays.  According to our rough predictions, many decay modes could be observed at BESIII,  LHCb, and Belle II, and some decay modes  might be measured in near future experiments. Therefore, the SU(3) flavor symmetry will be further tested by these semileptonic decays in future experiments.

\section*{ACKNOWLEDGEMENTS}
The work was supported by the National Natural Science Foundation of China, No. 12175088.

\section*{References}

\end{document}